\documentclass{aa}

\usepackage[T1]{fontenc}

\usepackage{balance}
\usepackage{txfonts}
\usepackage{orcidlink}

\usepackage{aecompl}
\usepackage{graphicx}
\usepackage{amsmath}
\usepackage{amssymb}

\usepackage{newtxtext,newtxmath}
\usepackage{pifont}

\usepackage{subfig}
\usepackage{comment}
\usepackage{multirow}
\usepackage{colortbl}
\usepackage{mathrsfs}
\usepackage{tabularx}
\usepackage{arydshln}
\usepackage{multirow}
\usepackage{graphicx}
\usepackage{txfonts}

\hypersetup{
    colorlinks=true,
    linkcolor=blue,
    filecolor=magenta,
    urlcolor=blue,
    pdftitle={Overleaf Example},
    pdfpagemode=FullScreen,
    citecolor=blue
    }

\usepackage{pifont}

\newcommand{\quotes}[1]{``#1''}
\newcommand\code[1]{\textsc{\MakeLowercase{#1}}}

\def\be{\begin{equation}}
\def\ee{\end{equation}}

\def\msun{{\rm M}_{\odot}}

\def\msunpc2{\msun/{\rm pc}^{2}}

\def\gsim{\lower.5ex\hbox{\gtsima}}
\def\lsim{\lower.5ex\hbox{\ltsima}}
\def\gtsima{$\; \buildrel > \over \sim \;$}
\def\ltsima{$\; \buildrel < \over \sim \;$} \def\gsim{\lower.5ex\hbox{\gtsima}}
\def\lsim{\lower.5ex\hbox{\ltsima}}
\def\simgt{\lower.5ex\hbox{\gtsima}}
\def\simlt{\lower.5ex\hbox{\ltsima}}

\def\CII{\hbox{[C~$\rm II $]}}
\def\OIII{\hbox{[O~$\scriptstyle\rm III $]}}
\def\OI{\hbox{[O~$\scriptstyle\rm I $]}}
\def\OII{\hbox{[O~$\scriptstyle\rm II $]}}

\def\CIII{\hbox{C~$\scriptstyle\rm III $]}}

\definecolor{mkcolor}{HTML}{0f6c20}
\definecolor{gr}{HTML}{116921}
\definecolor{bluepred}{HTML}{1E90FF}

\newcommand{\gszz}{JADES-GS-z14-0}
\newcommand{\gsz}{GS-z14-0}
\newcommand{\gszt}{JADES-GS-z14-1}

\def\CII{\hbox{[C~$\scriptstyle\rm II $]~}}
\def\CIII{\hbox{C~$\scriptstyle\rm III $]~}}
\def\OII{\hbox{[O~$\scriptstyle\rm II $]~}}
\def\OIII{\hbox{[O~$\scriptstyle\rm III $]~}}

\def\HII{\hbox{H~$\scriptstyle\rm II\ $}}

\def\circularity{\epsilon_\mathrm{circ}}

\begin{document}

\title{Amaryllis: A digital twin of the earliest galaxies in the Universe}
\titlerunning{Amaryllis: a digital twin of the earliest galaxies in the Universe}

\author{
        Mahsa Kohandel$^{1}$\thanks{\email{mahsa.kohandel@sns.it}}\orcidlink{0000-0003-1041-7865},
        Andrea Pallottini$^{1,2}$\orcidlink{0000-0002-7129-5761} and
        Andrea Ferrara$^{1}$\orcidlink{0000-0002-9400-7312}
       }

\institute{
        Scuola Normale Superiore, Piazza dei Cavalieri 7, I-56126 Pisa, Italy
        \and Dipartimento di Fisica ``Enrico Fermi'', Universit\'{a} di Pisa, Largo Bruno Pontecorvo 3, I-56127 Pisa, Italy
        }
\authorrunning{M. Kohandel et al.}

\date{Received May 12, 2025; accepted September 8, 2025}

\abstract
{
Synergies between JWST and ALMA are unveiling a population of bright, super-early galaxies  ($z>10$), including systems like GS-z14-0 ($z=14.2$) and GHZ2 ($z=12.3$) with extreme far-infrared (FIR) line ratios (${\rm [OIII]88\mu m/[CII]158\mu m} > 3$) that challenge galaxy formation models.
To address this, we identified a synthetic analog of these sources, \quotes{Amaryllis,} within the SERRA zoom-in simulations and tracked its evolution from  $z=16$ to $z=7$. During this period, Amaryllis grows from $\log(M_\star/M_\odot) \sim 7.4$ to $10.3$, linking super-early progenitors to the massive galaxy population at the end of reionization.
At $z \sim 11.3$, Amaryllis closely matches the observed properties of GS-z14-0, including the $M_\star$, star formation rate, and the luminosity of FIR ([OIII]~88$\mu$m) and UV (e.g., CIII]$1908$) lines.
We find high [OIII]/[CII] ratios during short, merger-driven starburst episodes, when low metallicity ($Z \sim 0.1\,Z_\odot$) and high ionization conditions ($U_{\rm ion} \sim 0.3$) push the interstellar medium far from equilibrium. These extreme FIR line ratios are thus transient and linked to major mergers that ignite strong ionized gas outflows.
Strikingly, despite this dynamical activity, Amaryllis develops a rotation-supported gaseous disk ($V/\sigma \sim 4$–6) by $z \sim 11$, while stars remain dispersion-dominated. This coexistence of ordered gas rotation and merger-driven disturbances occurs within a massive yet typical $\Lambda$ cold dark matter halo, enabling disk formation even at cosmic dawn.
}
\keywords{
          Galaxies: high-redshift --
          Galaxies: kinematics and dynamics -- Galaxies: structure -- Galaxies: evolution
         }

\maketitle

\section{Introduction}

The \textit{James Webb} Space Telescope (JWST) has led to extraordinary advancements in the study of galaxy formation, particularly at the highest redshifts ($z>10$) during cosmic dawn \citep[see][for a review]{Stark+25}. This previously uncharted epoch marks the emergence of the first stars and galaxies, offering an unprecedented window into early cosmic history.  JWST has already uncovered a substantial population of massive ($M_\star > 10^{8}M_\odot$) super-early ($z>10$) galaxies \citep{Arrabal23, atek:2023, Bunker23, Carniani+24z14nirspec, Castellano24, Curtis-Lake+23, Finkelstein23,Harikane+24_CEERS, Hsiao23, Wang23, Robertson23, Robertson24, Tacchella23, Zavala+25}, providing new insights into their assembly histories, star formation activity, and interstellar medium (ISM) conditions.

Building on these discoveries, the Atacama Large Millimeter/Submillimeter Array (ALMA) has been used to investigate fine-structure far-infrared (FIR) cooling lines, such as \OIII~$88\,\mu\rm{m}$ and \CII~$158\,\mu\rm{m}$ \citep{Bakx22, Fujimoto22, Kaasinen22, Popping22, Yoon22, Schouws+25}. However, early ALMA follow-ups often resulted in non-detection, particularly for GHZ2 \citep{Castellano22, Naidu22, Donnan22, Harikane22} and GHZ1 \citep{Treu22, Santini22}. These outcomes were initially attributed to sensitivity limits and the high cost of spectral scans \citep{Bakx22, Kaasinen22, Furlanetto22} or, in some cases like HD1 (originally at $z=13.3$), to misidentifications of low-redshift interlopers at $z\sim4$ \citep{Harikane+24}.

Alongside these observational efforts, theoretical studies using zoom-in cosmological simulations \citep{Kohandel+23, Nakazato+23} have investigated the detectability of FIR lines at such high redshifts. For instance, \citet{Pallottini+22} predict a large scatter in \OIII~$88\,\mu\mathrm{m}$ luminosity at fixed star formation rates (SFRs) and stellar mass in simulated galaxies. At $z>11$, \citet{Kohandel+23} find that many such galaxies lie slightly below the local metal-poor $L_{\rm{[OIII]}}$–SFR relation, driven by variations in the ionization parameter ($U_{\rm ion}$) regulating \OIII~emission \citep[see also][]{vallini:2017, Harikane22}. These ionizing conditions appear closely tied to morphological and dynamical states: \citet{Pallottini+22} show that $U_{\rm ion}$ correlates with merger stages, galaxy compactness, and \HII region sizes. Furthermore, \citet{Kohandel+23} find that transitioning from a clumpy, highly pressurized structure to a more extended, disk-like configuration, where the ionized ISM becomes more diffuse, can significantly boost \OIII~emission. Crucially, they predict that detecting  \OIII~$88\,\mu\mathrm{m}$~emission from $\sim 10^9\,M_\odot$ galaxies at $z>11$ could require as little as 2.8 hours of ALMA on-source time, provided those galaxies are already dynamically mature and exhibit a disk-like structure.

Such theoretical predictions are supported by recent JWST-ALMA synergies, which have led to the detection of \OIII~$88\,\mu$m emission in two galaxies at $z > 10$: \gszz~ (hereafter \gsz; at $z=14.1796$, with a 2.8-hour integration time; \citealt{Carniani+24z14alma, Schouws+24}), the most distant galaxy with a confirmed spectroscopic redshift, and GHZ2 (at $z=12.3327$; \citealt{Zavala+24}). Both galaxies lie on or slightly below the $L_{\rm[OIII]}$-SFR relation, consistent with the predictions from \code{SERRA} \citep{Kohandel+23}, and show substantial metal enrichment, with gas-phase metallicities estimated at $Z \sim 0.05-0.2 \rm Z_\odot$.
Most recently, \citet{Perez-gonzalez+25} report six F200W and three F277W dropout candidates, at $z\sim 17$ and $z\sim 25$, respectively, which, if confirmed, would seriously challenge standard $\Lambda$ cold dark matter ($\Lambda$CDM)-based galaxy formation models \citep{Matteri+25}.

We used the \code{SERRA} simulations to delve more deeply into the assembly history of these super-early galaxies. Our primary aim is to identify and characterize their progenitors, shedding light on their star formation, kinematics, and ISM conditions. By tracing how these galaxies form and evolve over cosmic time, we examined how early in time zoom-in cosmological simulations are able to track such nascent systems, and we assessed whether they could plausibly grow into the massive galaxies observed at $z\sim 6-7$.  Section \ref{sec:serra} provides an overview of the \code{SERRA} simulations, followed by the introduction of Amaryllis, a digital twin\footnotemark{} of observed $z>10$ galaxies, in Sect. \ref{sec:the_clone}. In Sect. \ref{sec:FIRdiversity} we examine the evolution of FIR line ratios, while Sect. \ref{sec:dynamics} explores the role of galaxy kinematics in the early Universe. Finally, our conclusions are summarized in Sect. \ref{sec:conclusions}.
\footnotetext{By ``digital twin,'' we mean a simulation that not only reproduces the
global properties of a real system, but also mimics its internal structure, dynamics,
and evolution in a physically motivated way. The term highlights the close physical
analogy between the simulated galaxy and its observed counterparts.}
\section{Summary of the SERRA simulations}\label{sec:serra}

The \code{SERRA} suite of zoom-in simulations \citep{Pallottini+22} is tailored to study the formation and evolution of early galaxies, from the cosmic dawn epoch ($z>10$) to the post-reionization epoch ($z\sim4$).
These simulations account for the interaction of radiation, gas, stars, and dark matter (DM) using the adaptive mesh refinement code \code{RAMSES} \citep{Teyssier+02, Rosdahl+15}.
Through the zoom-in technique, \code{SERRA} follows the formation of galaxies from intergalactic down to molecular cloud scales ($\simeq 30$ pc at $z\sim 6$), which is required to capture small-scale processes such as star formation and feedback within a cosmological context.
The model includes chemical nonequilibrium evolution \citep{Grassi+14,pallottini:2017althaea}, stellar feedback \citep{pallottini:2017dahlia}, and radiative transfer \citep{decataldo:2019,pallottini:2019}.

The \code{SERRA} project has made significant progress in bridging the gap between simulations and observations by accurately reproducing galaxy properties for direct comparison with current and future observational data \citep{behrens:2018dust, gelli+2020, gelli:2021, gelli:2023, Gelli+25, Kohandel+19, Kohandel+20, Kohandel+23, Kohandel+24, pallottini:2019, Pallottini+22, Pallottini+24, Pallottini+23, Rizzo+22, vallini:2018, vallini:2020, Zanella+21}. A key feature is the ability to calculate both line and continuum emission across multiple wavelengths.
Line emission is computed using the \code{CLOUDY} spectral synthesis code \citep{Ferland+17}, accounting for the turbulent structure of molecular clouds \citep{vallini:2018,pallottini:2019}. The attenuated UV and FIR continuum emission is modeled through the Monte Carlo radiative transfer code \code{SKIRT} \citep{Baes2015, Camps2015}, with dust properties implemented in post-processing \citep{behrens:2018dust}.

Moreover, \code{SERRA} can generate hyperspectral data cubes for key emission lines \citep{Kohandel+20}. These data cubes facilitate detailed spatial and kinematic analysis, capturing emission line profiles and spatial variations, while integrated line strengths enable a direct comparison with observational data \citep{Kohandel+19, Kohandel+24, Rizzo+22}. By modeling both global and spatially resolved emission, \code{SERRA} can provide insights into the physical conditions of the ISM in high-redshift galaxies, ensuring that the simulations are well suited for comparison with a wide range of observational datasets.
\begin{figure*}[t]
    \centering
    \includegraphics[width=0.95\textwidth]{amaryllis_sfh_and_merger_tree_and_composite_image.pdf}
    \caption[Comprehensive view of Amaryllis in evolution]{
        Overview of \quotes{Amaryllis,} a digital twin of early galaxies.
        \textit{Upper-left panel}: Merger history of Amaryllis. Each curve shows the stellar mass ($M_\star$) of the progenitors as a function of cosmic time ($t$), with the corresponding redshift ($z$) on the upper axis. The black line indicates the main galaxy, while each colored line represents a merging satellite. Solid lines mark the phases during which each system remains distinct, and dashed lines highlight the mergers. The observational measurements shown include massive $z>10$ galaxies, spectroscopically confirmed by JWST \citep[violet triangles;][Table 1]{Ferrara+24}, $z\sim6-9$ galaxies from CEERS and COSMOS \citep[green triangles;][]{Harikane+24_CEERS}, and REBELS sources \citep[red triangles;][Table 2]{Dayal+22}.
        \textit{Upper-right panel}: SFH of Amaryllis throughout cosmic time, with vertical lines corresponding to minor (dotted) and major (dashed) merger events.
        \textit{Bottom panels}: Composite images at four evolutionary stages, built from the stellar surface density ($\Sigma_\star$), gas surface density ($\Sigma_{\rm{gas}}$), and the Habing field intensity (G).
        \label{fig:amaryllis_comp}
        }
\end{figure*}

\section{Amaryllis: A digital twin of early galaxies}\label{sec:the_clone}

In \citet{Kohandel+23}, we identified a sample of 366 super-early ($11 \lesssim z \lesssim 14$) \code{SERRA} galaxies with $M_\star \gtrsim 10^8 M_\odot$.
From this sample, a subset of 42 galaxies was shown to have $6 \leq \text{SFR}/M_\odot\, \text{yr}^{-1} \leq 35$, i.e., to match the SFR values observed for super-early candidates by JWST.
Galaxies in such a subset reside in DM halos with masses $10^{10-10.7} M_\odot$ and are classified as starbursts based on their position in the $\Sigma_{\rm{SFR}} - \Sigma_{\rm{gas}}$ plane\footnote{The burstiness parameter, $\kappa_s = 10^{12} \Sigma_{\rm{SFR}}/\Sigma_{\rm{gas}}^{1.4}$ \citep{Ferrara:2019}, for these galaxies spans the range $\kappa_s \sim 2 - 150$ \citep{Pallottini+22}, indicative of a bursty star formation mode.}. On the $L_{\rm [OIII]}$-SFR relation, they lie between local metal-poor and local starburst galaxies \citep{DeLooze+14}, with ten objects slightly below the metal-poor relation, similar to \gsz\ and GHZ2 \citep[see][Fig. 1]{Kohandel+23}.

\subsection{A possible assembly history of super-early galaxies}

In this work we focused on \quotes{Amaryllis,} the brightest galaxy in that subgroup and representative of that below-relation population. To reconstruct its assembly history, we applied the \code{rockstar-galaxies} phase-space clustering algorithm \citep{behroozi:2013} and tracked its merger history with an approach similar to \code{MERGERTREE} \citep[][see Sect.~2.4.1 of \citealt{Pallottini+22} for details]{knebe:2013}, using time steps of $\Delta \simeq 10\,\mathrm{Myr}$. This method reliably follows the galaxy down to $z \sim 12$. At higher redshifts, however, the halo finder becomes unreliable due to the small sizes and irregular dynamical states of early structures. To extend the reconstruction to earlier epochs, we adopted a complementary method: stellar particles associated with the galaxy at lower redshift are tracked back across successive snapshots by matching their unique IDs directly within the simulation outputs. By centering the analysis on the stellar mass distribution and extracting local properties independently of halo catalogs, we can follow the galaxy's growth down to $z \sim 16$, corresponding to its initial formation in the simulation. Figure~\ref{fig:amaryllis_comp} provides a broad overview of Amaryllis, including
\textit{(i)} its merger history,
\textit{(ii)} star formation history (SFH), and
\textit{(iii)} composite images at four key epochs ($z=14, 11.3, 10.6,$ and $7.0$).

We begin with the merger history of Amaryllis, shown in the upper-left panel of Fig.~\ref{fig:amaryllis_comp}, which traces the stellar mass growth of the main progenitor and its merging companions from $z=16$ to $z=7$. Starting from $\log(M_\star/M_\odot)=7.36$ at $z=16$, Amaryllis assembles rapidly through a combination of cosmological accretion and hierarchical merging, reaching $\log(M_\star/M_\odot)=10.26$ by $z=7$. Steep rises in the mass-growth curve correspond to major mergers, which deposit stellar mass abruptly, while plateaus indicate phases of in situ star formation and gradual gas accretion. These episodes are also associated with sharp bursts in the SFR (see the SFH in the upper-right panel of Fig.~\ref{fig:amaryllis_comp}). At $z<12$, when our \code{rockstar}-based halo-finder is reliable, three major mergers occur at $t = 429$, $519$, and $567\,\mathrm{Myr}$, with the first playing a central role in the evolution discussed in the following sections.

The mass evolution of Amaryllis broadly aligns with that of spectroscopically confirmed galaxies at $z>10$. By $z\sim14$, its stellar mass of $\log(M_\star/M_\odot) \sim 8$ matches the range reported for \gsz~and \gszt~\citep{Carniani+24z14nirspec}. By $z\sim10$, it grows to $\log(M_\star/M_\odot) \sim 9.4$, consistent with other bright candidates such as CEERS2-588 \citep{Harikane+24_CEERS}. Its merging companions at $z>10$ have masses comparable to observed galaxies like GN-z11 and UNCOVER-37126 \citep{Bunker23, Wang+23}. While direct comparisons are limited by uncertainties in spectral energy distribution-derived stellar masses \citep[see, e.g.,][]{Helton+25}, the overall agreement suggests that Amaryllis represents a plausible analog of the most massive galaxies currently known at these redshifts. It also stands out as one of the first simulated galaxies to reach such a high stellar mass at $z>10$, a regime still challenging to access in cosmological simulations due to volume and resolution constraints (see \citealt{yang+25} for FIRE and Illustris-TNG simulations).

The subsequent growth of Amaryllis toward the end of the epoch of reionization is also particularly interesting. By $z\sim 7$, it reaches $\log(M_\star/M_\odot) = 10.26$, placing it among the most massive star-forming galaxies known at these epochs. This mass is consistent with estimates for sources in the REBELS sample \citep{Buowens+21}, such as REBELS-25 before JWST observations \citep{Dayal+22}. However, more recent JWST/NIRSpec data suggest that REBELS galaxies have lower stellar masses than previously inferred \citep{Rowland+24}. If the most distant galaxy observed to date, \gsz, follows a growth trajectory similar to that of Amaryllis, forming before $z=16$ and accumulating $\sim 80\%$ of its final stellar mass in less than $50\,\mathrm{Myr}$, it could potentially evolve into an object comparable to, or even more massive than, REBELS-25, entering the rare population of the most massive galaxies known at $z\sim 7$.

The bottom panel of Fig.~\ref{fig:amaryllis_comp} shows face-on composite views of Amaryllis at four key stages, built from stellar surface density ($\Sigma_\star$), gas surface density ($\Sigma_{\mathrm{gas}}$), and the Habing field intensity ($G$). In terms of morphology, gas in Amaryllis shows a disk-like structure with emerging spiral features ($10^2 \lesssim n/\mathrm{cm}^3 \lesssim 10^3$) already at $z>11$, while stars remain centrally concentrated. The radiation field is clumpy, with intense peaks ($G > 10^3 G_0$, where $G_0=1.6 \times 10^{-3}\,\mathrm{erg}\,\mathrm{cm}^{-2}\,\mathrm{s}^{-1}$ is the Milky Way value) tracing recent star formation. As the system undergoes repeated mergers, particularly the first major one $t = 429\,\rm{Myr}$, it enters a more disturbed dynamical phase. By $z\sim7$, it settles into a well-defined rotating disk with spiral arms, marking the reestablishment of rotational support. In Sect.~\ref{sec:dynamics} we examine how these interactions shape the morpho-kinematic evolution and ISM structure in more detail.

\begin{figure}[t]
    \centering
    \includegraphics[width=0.5\textwidth]{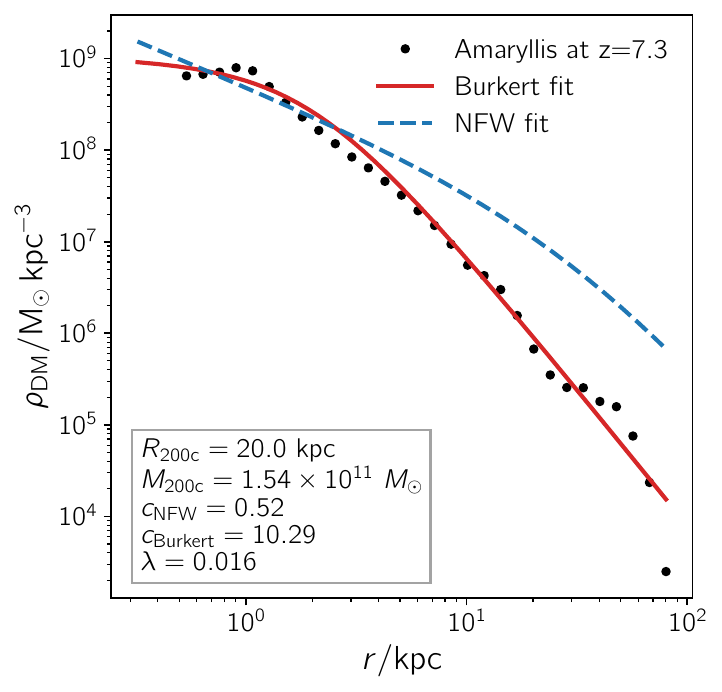}
    \caption{DM halo structure of Amaryllis at $z=7.3$.
    We show a spherically averaged DM density profile (black points) fitted with a Burkert (solid red) and an NFW (dashed blue) profile.
    \label{fig:amaryllis_halo}}
\end{figure}

\subsection{Dark matter halo properties\label{sec:halo}}
To place Amaryllis in the broader context of $\Lambda$ CDM, we studied the properties of its host DM halo at $z=7.3$, i.e., close to the redshift of REBELS–25 (the most distant galaxy with a measured rotation curve; \citealt{Rowland+24}). At this epoch, the halo encloses a mass of 
$M_{200\mathrm{c}} \simeq 1.5\times10^{11}\,M_\odot$ within a radius of $R_{200\mathrm{c}} \simeq 20\,\mathrm{kpc}$, as derived from the particle–based enclosed–density profile (Fig.~\ref{fig:amaryllis_halo}). The corresponding virial velocity is $V_{200\mathrm{c}}=\sqrt{GM_{200\mathrm{c}}/R_{200\mathrm{c}}}\approx200\,\mathrm{kms^{-1}}$. 
Such values place Amaryllis in the high-mass tail of the expected halo distribution \citep[e.g.,][]{behroozi:2013}. 
The angular–momentum content of the halo was quantified by the dimensionless spin parameter in the \citet{Bullock2001} form, 
\begin{equation}
\lambda = \frac{J}{\sqrt{2}\,M_{200\mathrm{c}}\,V_{200\mathrm{c}}\,R_{200\mathrm{c}}},
\end{equation}
where $J$ is the total angular momentum of DM particles within $R_{200\mathrm{c}}$. For Amaryllis we measure $\lambda \simeq 0.016$, a value on the low-to-typical side of the expected lognormal distribution 
($\langle\lambda\rangle\simeq0.03$, $\sigma_{\ln\lambda}\simeq0.5$; \citealt{Maccio2007, Dutton2014}), consistent with a system that has experienced recent mergers but is not an outlier in $\Lambda$CDM.
The internal density structure is best reproduced by a cored Burkert profile \citep{Burkert1995},
\begin{equation}
\rho_\mathrm{Bur}(r)= \frac{\rho_0}{(1+r/r_c)\,[1+(r/r_c)^2]}\,,
\end{equation}
with a core radius $r_c\simeq1.9\,\mathrm{kpc}$, corresponding to a formal concentration $c_{\rm Bur} \equiv R_{200\mathrm{c}}/r_c \simeq 10.3$, which we use here only as a proxy for halo compactness, not as a direct analog of NFW concentration.
By contrast, an NFW fit \citep{Navarro1997} produces systematically higher densities in the central few kiloparsecs, leaving significant residuals relative to the simulation. Such deviations from cuspy profiles are a well-known outcome of bursty stellar feedback, which drives central potential fluctuations that flatten the cusp into a core \citep[e.g.,][]{Pontzen2012, Lazar2020}.

These properties indicate that Amaryllis resides in a massive yet not unusual $\Lambda$CDM halo. The combination of modest spin and relatively high concentration creates a potential that is deep enough to confine baryons but dynamically responsive to feedback and mergers.
This environment naturally explains the baryon–dominated, rotation–supported gaseous disk predicted in massive ($M_\star>10^{10}M_\odot$) epoch of reionization galaxies \citep{Kohandel+24} and described for Amaryllis in Sect.~\ref{sec:dynamics}.

\subsection{Emission line properties}

\begin{table*}[t]
\caption{Properties of Amaryllis compared with those of \gsz~ and GHZ2.}
\renewcommand{\arraystretch}{1.35}
\centering
\setlength{\tabcolsep}{12pt}
\begin{tabular}{|l|c|c|c|c|}

\hline
\rowcolor{gray!30} \textbf{Property} & \textit{\textbf{Amaryllis}} & \textbf{\gsz} & \textbf{GHZ2} & \textbf{Units} \\
\hline
\rowcolor{gray!10} \textbf{General} & & & & \\
Stellar mass ($M_{\star}$) & $8.8 \times 10^{8}$ & $\,4.0^{+15}_{-1.5}\times 10^{8}\,^a$ &  $8.1^{+1.9}_{-3.9} \times 10^{8}\,^d$ & $M_\odot$ \\
Gas mass ($M_{g}$) & $4.5 \times 10^{8}$ & $4.0\,^a - 63\,^k \times 10^{8}$ & - & $M_\odot$ \\
Dust mass ($M_{d}$)& $1.2 \times 10^{5}$ & $1.2^{+0.3}_{-0.3}\times 10^{4}\,^i$ & $6.1^{+10.6}_{-6.1}\times 10^{4}\,^i$ & $M_\odot$ \\
Halo virial mass ($M_{h}$)& $5.1 \times 10^{10}$ &- & - & $M_\odot$ \\
Gas fraction ($f_g$) & $32\% $ & $\sim 30^a-90^k\%$ & - & - \\
    Gas phase metallicity (Z) & $0.2^*$ & $0.05-0.2\,^b$ & $0.05-0.1\,^f$ & $Z_\odot$ \\
SFR$_{10 \rm Myr}$ & $18$ & $19\pm 6\,^a$ &  $5.2^{+1.1}_{-0.6}\,^d$ & $M_\odot \, \text{yr}^{-1}$ \\
Stellar Age & $42.9$ & $40\pm 5\,^b$ & $26\pm 55\,^h$ & Myr \\
    Gas number density ($n$) & $810^*$ & $51^{+116}_{-32}\,^j$ & - & $\rm{cm}^{-3}$ \\
    Ionization parameter ($U_{\rm{ion}}$) & $0.01^*$  & $0.004^{+0.003}_{-0.001}\,^b$ & $0.02^{+0.01}_{-0.01}\,^d$ &  \\
\hline

\rowcolor{gray!10} \textbf{UV and Optical Emission Lines} & & & & \\
\CIII~$\lambda$1908  & $3.0\times10^{8}$ & $1.1^{+0.3}_{-0.3}\times10^{8}\,^a$ & $5.14^{+0.11}_{-0.11}\times10^{8}\,^f$ & L$_\odot$ \\
    \OIII~$\lambda$5007 & $2.1\times10^{9}$ & - & - & L$_\odot$ \\
    \OII~$\lambda$3727 & $4.3\times10^{8}$ & - & $1.52^{+0.63}_{-0.63}\times10^{8}\,^f$ & L$_\odot$ \\
    H$\alpha$~$\lambda$6563 & $3.4\times 10^{9}$ & - & - & L$_\odot$ \\
    H$\beta$~$\lambda$4861 &$1.1\times 10^{9}$ & - & - & L$_\odot$ \\
H$\alpha$~ half light radius & 230 & - & - & pc \\
H$\beta$~ half light radius & 227 & - & - & pc \\
\hline
\rowcolor{gray!10} \textbf{FIR Emission Lines} & & & & \\
\CII~$158\mu$m & $5.4\times10^{7}$ & $< 6\times10^{7}\,^j$ & - & L$_\odot$ \\
    \OIII~$88\mu$m & $1.2\times10^{8}$ & $2.0^{+0.5}_{0.4}\times10^{8}\,^{b,c}$ & $1.7^{+0.4}_{0.4}\times10^{8}\,^e$ & L$_\odot$\\
\OI~$63\mu$m & $4.9\times10^{7}$ & - & - & L$_\odot$ \\
\OIII~$52\mu$m & $2.1\times10^{8}$ & - & $< 9.6\times10^{8}\,^e$ & L$_\odot$ \\
\CII~ half light radius & 209 & - & - & pc \\
\OIII~ half light radius & 170 & - & - & pc \\
\OI~ half light radius & 169 & - & - & pc \\
\rowcolor{gray!10} \textbf{Continuum UV and FIR} & & & & \\
$M_{\rm UV}$ & -18.98 & $-20.81\pm 0.16\,^a$& -20.53$\pm 0.01\,^d$& - \\
UV half light radius & 63 & $260^{+2}_{-2}\,^a$ & $50 \,^h$& pc \\
A$_V$ & 3.3 & $0.3^{0.14}_{0.07}\,^a$ &  $0.04^{0.07}_{0.03}\,^d$ & - \\
$L_{\rm IR}$ & 9.1$\times 10^{10}$ & $<1.3\times10^{11}\,^a$ & - & L$_\odot$ \\
IR half light radius & 34 & - & - & pc \\
\hline
\end{tabular}
    \tablefoot{Values for \gsz~and GHZ2 compiled from the literature: $^a$\citet{Carniani+24z14nirspec}, $^b$\citet{Carniani+24z14alma},$^c$\citet{Schouws+24}, $^d$\citet{Castellano+24}, $^e$\citet{Zavala+24}, $^f$\citet{Calabro+24}, $^h$\citet{Zavala+25}, $^i$\citet{Ferrara+24},$^j$\citet{Schouws+25}, $^k$\citet{Heintz+25},$^*$ These properties are averaged within the galaxy’s half–gas-mass radius ($\sim 169\rm{pc}$).}
\label{tab:properties}
\end{table*}

\begin{figure}[t]
    \centering
    \includegraphics[width=0.5\textwidth]{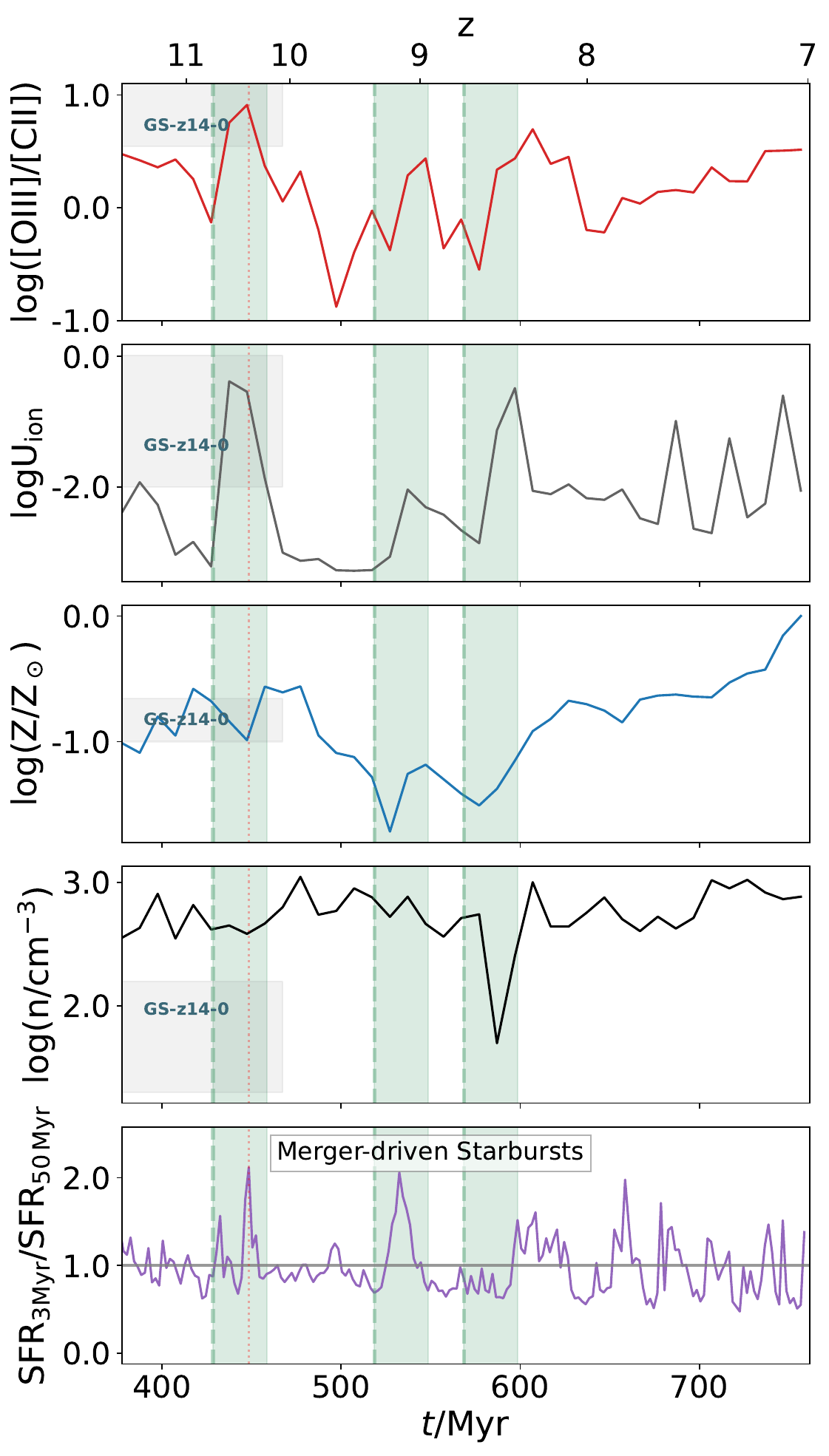}
    \caption{Temporal evolution of various Amaryllis properties:
    $L_{\rm{OIII}}/L_{\rm{CII}}$ (\textit{top panel}) , the ionization parameter $U_{\rm{ion}}$ (\textit{second panel}), gas metallicity, $Z$ (\textit{third panel}), gas number density, $n$ (\textit{fourth panel}), and star-formation variability, $\rm{SFR}_{3\,\rm{Myr}}/\rm{SFR}_{50\,\rm{Myr}}$ (\textit{bottom panel}).
    The correlations between $L_{\rm{OIII}}/L_{\rm{CII}}$ and each other property are presented in Table \ref{tab:correlation}. The vertical dashed green lines correspond to major merger events, and the dotted red line indicates the snapshot with the highest [OIII]/[CII].
    \label{fig:oiii_to_cii_evolution}}
\end{figure}

\begin{table}[t]
\caption{Correlations of [OIII]/[CII] in Amaryllis.}
\centering
\resizebox{0.5\textwidth}{!}{
\begin{tabular}{lcccc}
\hline
 & $\log U_{\mathrm{ion}}$
 & $\log \tfrac{\mathrm{SFR}_{3\,\mathrm{Myr}}}{\mathrm{SFR}_{50\,\mathrm{Myr}}}$
 & $\log \tfrac{\rm n}{\mathrm{cm}^{-3}}$
 & $\log \tfrac{Z}{Z_{\odot}}$ \\
\hline
    $r_s$ & $0.64$ & $0.39$ & $-0.11$ & $0.26$\\
    $p$   & $1.4\times 10^{-5}$ & $1.5\times 10^{-2}$ & $5\times 10^{-1}$ & $1.1\times 10^{-2}$\\
\hline
\end{tabular}
}%
\tablefoot{Spearman correlation coefficients ($r_s$) and $p$-values for the relation between [OIII]/[CII] and the selected physical properties.}
\label{tab:correlation}
\end{table}

To place Amaryllis in the context of recent ALMA detections of $z>10$ galaxies \citep{Carniani+24z14alma, Schouws+24, Zavala+24}, we followed the evolution of its \OIII\,88\,$\mu$m luminosity from $z=16$ to $z=7$. Our main predictions rely on \code{CLOUDY}-based radiative transfer models \citep{Pallottini+22}, applicable at $z\lesssim12$ where the ISM is sufficiently resolved.
At earlier times, we adopted a simplified, physically motivated model that assigns \OIII~emission based on local gas properties (density, temperature, and metallicity) and the global SFR (see Appendix~\ref{sec:appendix1}).
This approximation is required because at this redshift the stellar mass is only $M_\star \sim 10^7\,\msun$, so a mass resolution of $\sim 10^4\,\msun$ combined with highly flickering star formation \citep{Pallottini+23} leads to a noisy spatial distribution of young stars. In addition, at high gas densities and with a spatial resolution of $\simeq 10\,\rm{pc}$, ionization fronts are not numerically resolved \citep{pallottini:2019, decataldo:2019}. The combination of these limitations hampers the ability to make accurate spatially resolved predictions for emission lines from highly ionized species, such as [OIII], which originate from O$^{++}$ and require a high ionization potential. Nevertheless, this approximation captures the expected scaling relations in unresolved star-forming regions and yields $L_{\mathrm{[OIII]}}\sim10^8\,L_\odot$ at $z\sim14$.

For a robust comparison with observed galaxies, we selected a representative snapshot at $z\sim11.3$, where the ISM is well resolved and line emission predictions are reliable. This stage is shown in the bottom panel of Fig.~\ref{fig:amaryllis_comp}. Table~\ref{tab:properties} summarizes the physical and emission-line properties of Amaryllis at this epoch, alongside those of two spectroscopically confirmed galaxies with ALMA+JWST coverage, \gsz\ and GHZ2. Included quantities span bulk properties, UV/optical lines (\CIII~$\lambda$1908, \OIII$\lambda$5007, \OII$\lambda$3727, H$\alpha$, H$\beta$), FIR lines (\CII~$158\,\mu$m, \OI~$63\,\mu$m, \OIII~$88\,\mu$m), and continuum emission.

Amaryllis shares many characteristics with observed super-early systems. Its gas fraction ($f_g \sim 32\%$), average number density ($n \sim 136\,\mathrm{cm}^{-3}$), and metallicity ($Z \sim 0.03\,Z_\odot$) are consistent with values inferred for both \gsz\ and GHZ2 \citep{Carniani+24z14nirspec}. The SFR at this stage is $\sim18\,M_\odot\,\mathrm{yr}^{-1}$, closely matching \gsz. Its FIR line luminosities are also comparable: \OIII~$88\,\mu$m falls within the range observed in both targets, while \CII~$158\,\mu$m remains consistent with the upper limit for \gsz\ from recent ALMA non-detections \citep{Schouws+25}.

There are, however, relevant differences in the UV and FIR continuum properties. Amaryllis has a fainter UV magnitude ($M_{\rm UV} \approx -18.98$) and a more compact UV half-light radius ($63\,\rm{pc}$), compared to \gsz\ ($260\,\rm{pc}$) but comparable to GHZ2 ($50\,\rm{pc}$). Its compact morphology, combined with high attenuation ($A_V \sim 3.3$), suggests that Amaryllis, like other \code{SERRA} galaxies \citep{Pallottini+22}, is caught in an obscured phase of star formation.
Part of these differences could be due to the limited modeling of pre-supernova feedback and radiation pressure invoked by the attenuation-free model \citep{Ferrara+23, Ferrara24a} to clear the dust and to drive mini-quenching episodes \citep{gelli:2023, Gelli+25}, as well as the lack of turbulence-informed post-processing for continuum emission \citep{dimascia:2025}.
Moreover, assumptions such as a fixed dust-to-metal ratio and a Milky Way-like extinction curve may not fully capture the diversity of dust properties and ISM conditions in high-redshift galaxies \citep{Markov+23, Markov+24, fisher:2025}. Addressing these limitations is beyond the scope of this paper, but it emphasizes the need for refined dust modeling in future work.

\section{Drivers of the \texorpdfstring{[OIII]/[CII]}{} line ratio} \label{sec:FIRdiversity}

As ALMA continues to detect multiple FIR lines in $z>10$ galaxies, it becomes possible to apply ISM diagnostics developed at lower redshifts to the earliest cosmic epochs. Among these, the [OIII]\,$88\,\mu$m / [CII]\,$158\,\mu$m luminosity ratio ([OIII]/[CII]) stands out as a powerful probe of the ionization state and structure of the ISM. At $z \sim 6$, this ratio is observed to span a broad range (1–20), significantly higher than in local starbursts or dwarfs \citep{Harikane+20, Carniani+20}.
In the case of \gsz, ALMA non-detection of [CII] yields a 3$\sigma$ upper limit of $L_{\mathrm{[CII]}} < 6 \times 10^7\,L_\odot$, implying [OIII]/[CII]$ > 3.5$ \citep{Schouws+25}. Photoionization modeling suggests this galaxy has a relatively low gas density ($n \sim 50\,\mathrm{cm}^{-3}$), high gas phase metallicity ($\sim 0.16\,L_\odot$) and extreme ionization conditions ($\log U_{\rm ion} > -2$). However, the origin of such elevated ratios—whether due to extreme metallicity, ISM conditions, or dynamical state—remains debated \citep{Arata+20, Pallottini+22}.

To investigate this further, we studied the evolution of [OIII]/[CII] in Amaryllis. As shown in the top panel of Fig.~\ref{fig:oiii_to_cii_evolution}, this ratio varies substantially over time, peaking at [OIII]/[CII] $\sim 8$ at $z \sim 10.4$, about $20\,\mathrm{Myr}$ after its first major merger. This high value corresponds to a phase of intense, merger-driven star formation. At this moment, the ISM is relatively metal-poor ($Z \sim 0.02\,Z_\odot$), strongly ionized ($U_{\rm ion} \sim 0.1$), and relatively low density ($n \sim 49\,\rm{cm}^{-3}$). Our density and ionization parameter are in line with the results of idealized photo-ionization models for \gsz~ in \citealt{Schouws+25}, although our gas-phase metallicity estimate is an order of magnitude lower.

Following this peak, the line ratio declines sharply (from 8 to 1 in within 40 Myr) in correspondence with a $U_{\rm ion}$ drop. This could suggest that strong outflows produced by starbursts disperse the central gas reservoir, reducing the ionization parameter and temporarily suppressing [OIII] emission. Subsequent gas inflows associated with later mergers replenish the ISM, reigniting star formation and restarting the cycle. Indeed, the peak in Amaryllis at $z \sim 10.4$—shortly after its first major merger—is accompanied by broad \OIII~line wings (see Sect.~\ref{sec:outflows}), hinting at powerful outflows.

To better understand the drivers of these variations, we correlated [OIII]/[CII] with the ionization parameter ($U_{\rm ion}$), gas-phase metallicity ($Z$), gas density ($n$), and star formation variability defined as $\mathrm{SFR}_{3\,\mathrm{Myr}} / \mathrm{SFR}_{50\,\mathrm{Myr}}$ \citep{Gelli+25, Endsley:2024}. As shown in the bottom panels of Fig.~\ref{fig:oiii_to_cii_evolution}, [OIII]/[CII] tracks both $U_{\rm ion}$ and SFR variability, peaking during recent bursts and in highly ionized phases. The Spearman coefficients listed in Table~\ref{tab:correlation} confirm that [OIII]/[CII] is most strongly correlated with $U_{\rm ion}$ ($r_s = 0.68$), moderately with SFR variability ($r_s = 0.39$), and weakly with $n$ ($r_s = 0.16$) and $Z$ ($r_s = 0.26$).

These findings suggest that elevated [OIII]/[CII] ratios at cosmic dawn primarily arise from bursty star formation episodes triggered by galaxy mergers. Such an interpretation naturally explains the observed properties of \gsz, which is luminous in [OIII] yet undetected in [CII]. If \gsz\ is indeed observed during a similar merger-induced burst phase, our models suggest that its gas-phase metallicity could be substantially lower ($Z\sim0.02\,Z_\odot$) than previously inferred from simplified photoionization modeling, and it may host ionized gas outflows. Nevertheless, a critical ambiguity remains: although merger-driven starbursts clearly enhance ionization, thereby elevating [OIII]/[CII], disentangling feedback-driven outflows from kinematic disturbances directly associated with ongoing mergers remains challenging. To address these complexities, we explored in detail the kinematic signatures of Amaryllis during this crucial post-merger phase.

\section{Galaxy kinematics at cosmic dawn}\label{sec:dynamics}

\begin{figure*}[t]
    \centering
    \includegraphics[width=\textwidth]{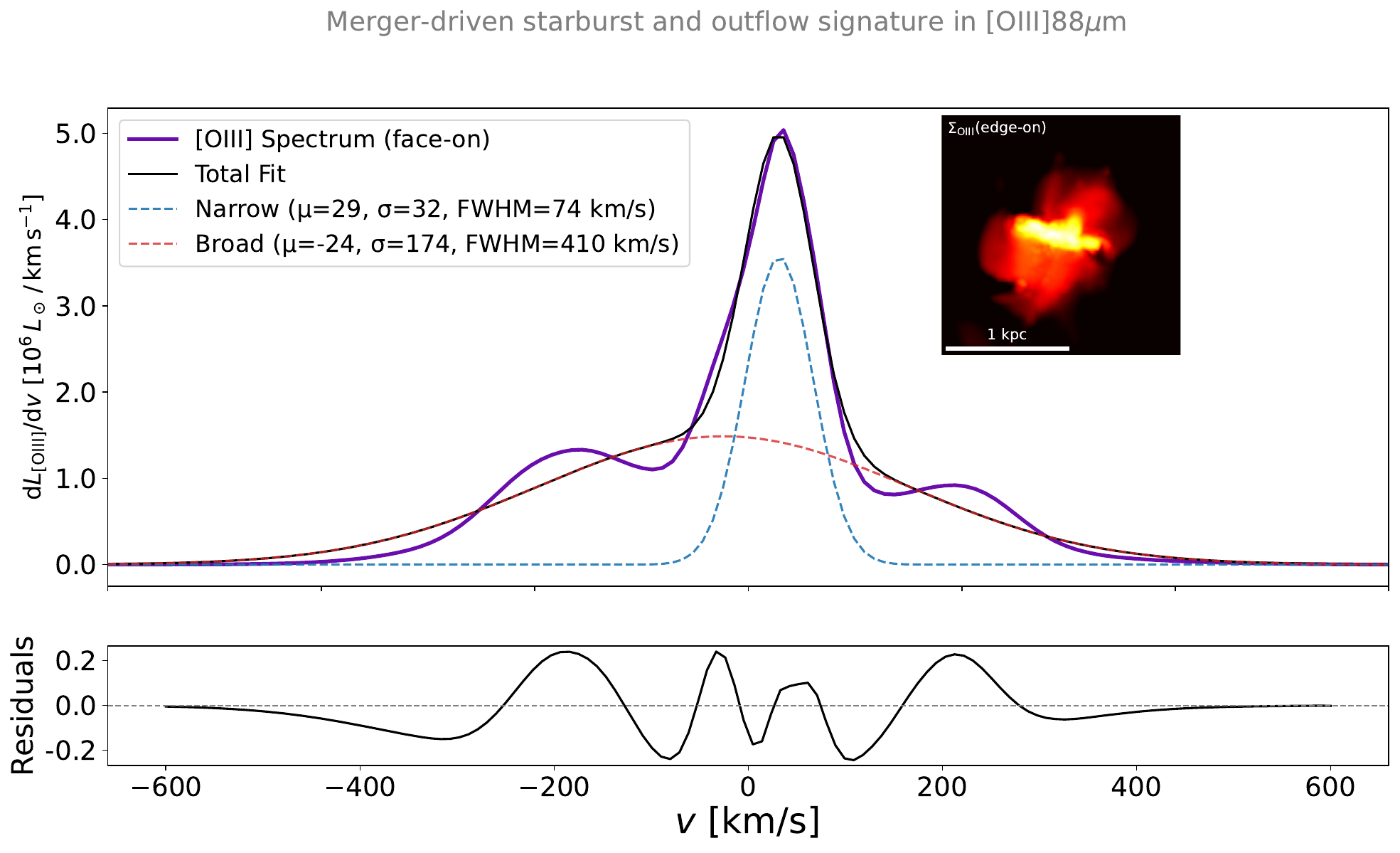}
    \caption[Merger-driven outflow signature in \OIII]{Integrated kinematic diagnostics of Amaryllis during its post-merger starburst phase at $z \sim 10.4$.
    \textit{Main panel}: Integrated \OIII~$88\,\mu$m spectrum from a face-on orientation, highlighting a prominent narrow component tracing star-forming gas near systemic velocity, and a broad, blueshifted wing indicative of ionized outflows. The overlaid two-Gaussian fit separates these components, with residuals shown in the bottom subplot. The inferred outflow velocity is $v_{\rm outflow} \simeq 258\,\mathrm{kms}^{-1}$. \textit{Inset panel}: Spatial distribution of [OIII] emission in the edge-on projection, revealing extra-planar ionized gas consistent with a large-scale outflow.
    \label{fig:oiii_outflow}}
\end{figure*}
\subsection{Outflows are interconnected with mergers}\label{sec:outflows}

To directly probe the kinematic impact of bursty star formation, we examined the \OIII~$88\,\mu$m line profile of Amaryllis during its peak post-merger phase at $z \sim 10.4$. As discussed in the previous section, this epoch coincides with a sharp rise in the [OIII]/[CII] ratio driven by highly ionized conditions following a major merger. Figure~\ref{fig:oiii_outflow} presents the integrated [OIII] spectrum from a face-on orientation, along with corresponding maps of the mean velocity and velocity dispersion. The [OIII] line profile is characterized by a prominent narrow core and broad, asymmetric wings extending to $\pm 400\,\mathrm{kms^{-1}}$. Such spectral features are traditionally interpreted as signatures of galactic-scale outflows, and the face-on viewing geometry adopted here minimizes contributions from rotational motions, thus enhancing the clarity of the outflow signature.

To quantitatively assess these kinematics, we fit the integrated [OIII] spectrum with two Gaussian components: a narrow core (FWHM $\simeq 74\,\mathrm{kms^{-1}}$), tracing star-forming gas at systemic velocity, and a broad, blueshifted component (FWHM $\simeq 410\,\mathrm{kms^{-1}}$) indicative of high-velocity ionized gas outflows. From these Gaussian fits, we estimated the outflow velocity using the widely adopted relation \citep{Rupke+05}
\begin{equation}
v_{\rm outflow} = |\mu_{\rm broad} - \mu_{\rm narrow}| + \frac{\mathrm{FWHM}_{\rm broad}}{2}\,,
\end{equation}
which yielded $v_{\rm outflow}\simeq258\,\mathrm{kms^{-1}}$. This ionized gas outflow velocity is higher than those observed in local galaxies with comparable stellar masses \citep{Marasco+23}, yet somewhat lower than the range found for low-mass star-forming galaxies at $3<z<9$ \citep{Carniani+24}.

However, we caution that the two-component Gaussian model leaves residual emission at $|v|\gtrsim 200\,\mathrm{kms^{-1}}$, suggesting the presence of additional, unresolved kinematic structures. These residuals may reflect complex outflow geometries, inflows, or merger-driven disturbances not fully captured by the model. Indeed, additional insight from our simulations reveals that the interpretation of the broad [OIII] wings is not straightforward. While the kinematic and morphological signatures are consistent with ionized outflows, a deeper analysis indicates that both inflowing and outflowing gas contribute significantly to the broad [OIII] wings, suggesting a complex, multiphase velocity field rather than a purely outflow. At $z = 10.4$, outflows and inflows have comparable specific kinetic energies ($\sim\!6\text{--}8 \times10^{13}\,\mathrm{erg\,g^{-1}}$). Given these limitations, we refrain from over-interpreting our results based on this simplified empirical approach. A detailed discussion of the energetics, metal loading, and multiphase velocity structure will be presented in a forthcoming paper.

At $z=10.4$, Amaryllis is still in the final stages of merger coalescence, with two distinct stellar and metallicity peaks separated by less than $100\,\mathrm{pc}$, despite its DM halo appearing dynamically relaxed. Such interactions naturally induce strong turbulent motions and may significantly contribute to the broad wings observed in the integrated [OIII] profile.

Crucially, the edge-on projection of [OIII] emission shown in Fig.~\ref{fig:oiii_outflow} reveals a clear extra-planar, ionized gas, further supporting the interpretation of outflows. In contrast, the cold gas traced by the [CII] line at the same epoch does not show comparable features, reinforcing the conclusion that the broad [OIII] emission primarily traces ionized outflows rather than rotation-supported components. Nevertheless, this case highlights the fundamental complexity intrinsic to high-redshift galaxy kinematics, where distinguishing purely feedback-driven outflows from dynamical merger signatures is particularly challenging.

In Amaryllis at $z=10.4$, we carefully minimized rotational contamination by aligning the galaxy face-on based on the full three-dimensional stellar velocity information. While such precise control is achievable in simulations, it is rarely attainable in observations. Thus, to robustly quantify outflows and distinguish them from merger-driven turbulence or rotation-induced velocity broadening at early cosmic epochs, we need alternative observational strategies and diagnostics. This consideration is especially critical if rotationally supported gaseous disks form earlier than previously thought, as suggested by recent observations and simulations at high z \citep[e.g.,][]{Rowland+24, Kohandel+24}. Therefore, in the next section we investigate when such disks might form in massive, early galaxies, and how their early emergence complicates interpretations of observed kinematic signatures.

\subsection{Early gas disk formation}

\begin{figure*}[t]
    \centering
    \includegraphics[width=0.95\textwidth]{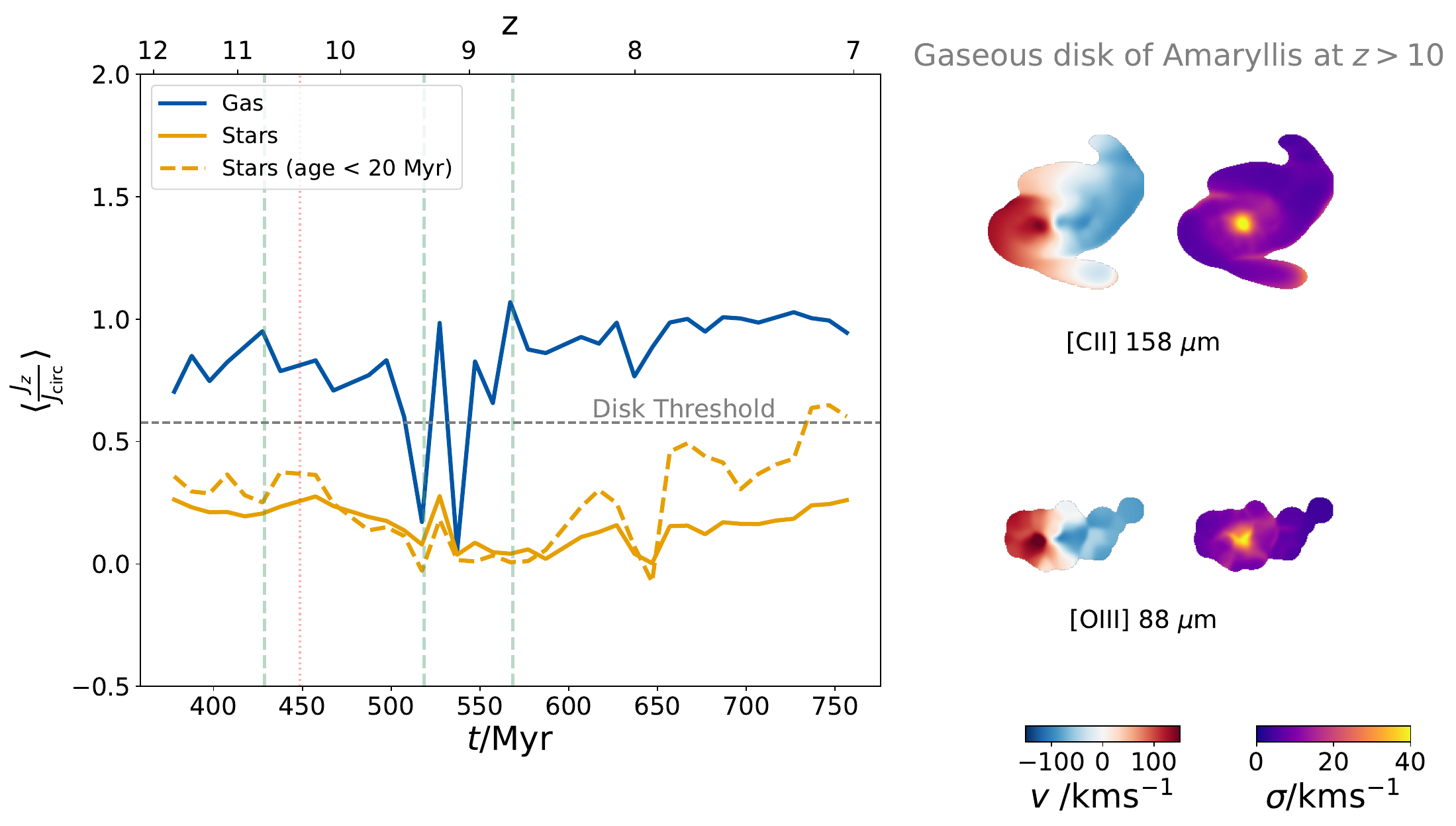}
    \caption[Gas and stellar dynamics of Amaryllis]{
    \textit{Left panel}: Evolution of mean circularity parameter, $\langle\circularity \rangle= \langle J_z/J_\mathrm{circ} \rangle$, with cosmic time ($t$) and corresponding redshift ($z$) for Amaryllis. The horizontal dashed line indicates the disk threshold ($1/\sqrt{3}$, \citealt{Simons+19}), while the shaded region highlights the triple merger phase.
    \textit{Right panels}: Velocity ($v$) and velocity dispersion ($\sigma$) maps of the gaseous disk in \CII~and \OIII~emission, showcasing the early gaseous disk properties at $z\simeq 11$ in FIR emission lines observable with ALMA at these redshifts.
    \label{fig:dynamics}
}
\end{figure*}

Recent high-resolution ALMA and JWST observations indicate that rotationally supported galaxy disks may have formed surprisingly early, even at cosmic dawn \citep{Rizzo+20, Roman-Oliveira+23, Rowland+24, Ferreira+22a, Kartaltepe+23, Kohandel+24}. To test this scenario, we quantified disk formation in Amaryllis using the circularity parameter ($\circularity\equiv J_z/J_{\mathrm{circ}} $; \citealt{Simons+19}) defined as the ratio between the gas or stellar angular momentum aligned with the galaxy's rotation axis ($J_z$) and the angular momentum of a circular orbit at the same radius ($J_{\mathrm{circ}}$)\footnotemark{}, and used in simulated galaxies \citep[e.g.,][]{Ceverino+15, Zana+22, Rizzo+22}. To implement this method, we identified the galactic center following \citet{Zana+22}, locating the minimum stellar gravitational potential via the tree algorithm of \citealt{Barnes+86}, as implemented by \citet{Grudic+21}. We then iteratively determined the galaxy's radial extent ($R_{\mathrm{eff}}$) and its vertical height, which isolates a coherent, rotating structure by progressively excluding inflowing or outflowing gas outside the vertical scale height.
Once $J_z$ and $J_{\mathrm{circ}}$ were computed for each resolution element, we derived the mass-weighted average circularity for the cold gas ($T\le 1.5\times 10^{3}$ K), all stellar particles, and the subset of stars younger than 20 Myr.

Following \citet{Simons+19}, components with $\langle \circularity \rangle > 1/\sqrt{3}$ ($< 1/\sqrt{3}$) were considered to  be rotation-supported (dispersion-dominated). In Fig.~\ref{fig:dynamics}, by tracing the circularity of gas and stars, we see that the cold gas in Amaryllis remains largely rotation-supported ($1/\sqrt{3} < \langle \circularity \rangle < 1$) throughout its evolution. However, we should be careful not to over-interpret these results, especially during the merger phase, as the algorithm might not be adequate for such complex dynamical structures.

Interestingly, the stellar component remains dispersion–dominated for most of the evolution: only by $z\sim 7$ do newly formed stars inherit the ordered motions of the gas disk, raising their circularity above the rotation–dominated threshold. Thus, cold gas settles into a disk structure nearly $300$ Myr before stars do. This offset reflects gas dissipation after mergers, while collisionless stars retain past dynamical memory. This is consistent with the halo properties (see Sect. \ref{sec:halo}). The modest spin ensures that baryons are not spread too diffusely, while the feedback–flattened inner profile lowers central shear and facilitates the regrowth of a cold gaseous disk after each merger. The resulting system is baryon–dominated in its inner few kiloparsecs, where gas rotation defines the dynamical state even though stars lag behind in angular–momentum buildup.

To connect with observations, the right panels of Fig.~\ref{fig:dynamics} feature mock kinematic maps of Amaryllis at $z\simeq11$ and $i=60^\circ$ in a non-merger phase at $z>10$. Both \CII\ and \OIII\ show a clear velocity gradient and centrally peaked dispersion, with $\sigma_{\rm[OIII]}\approx17\,\rm{kms}^{-1}$ and $\sigma_{\rm[CII]}\approx11\,\rm{kms}^{-1}$, yielding $V/\sigma\sim4$–6 depending on the emission line used. These values place Amaryllis firmly in the “cold disk” regime. In comparison, tentative ALMA measurements for \gsz\ report $\sigma_{\rm[OIII]}<40\,\rm{kms}^{-1}$ and $V/\sigma>2.5$ \citep{Scholtz+25}, consistent with our predictions. Deeper, higher‐resolution ALMA observations of \OIII\,$88\,\mu$m could thus reveal one of the most distant dynamically cold disks to date and help disentangle disk rotation from merger‐driven turbulence and outflows at $z>10$.

In short, Amaryllis demonstrates that dynamically cold gaseous disks can emerge in ordinary $\Lambda$CDM halos as early as $z>11$. Their early survival is enabled by baryon dissipation and a feedback–responsive halo potential, while stellar disks lag behind until fresh stars form in situ within the gas disk. This coexistence of ordered gas rotation and dispersion-dominated stars may be a defining feature of massive early galaxies.

\footnotetext{For each particle, $J_z$ is the component of the specific angular momentum aligned with the galaxy’s net angular momentum; $J_{\mathrm{circ}} = \sqrt{G M(<r)r}$ is the specific angular momentum of a notional circular orbit at radius $r$, assuming spherical symmetry for the enclosed mass $M(<r)$ (baryonic + DM), with $G$ being the gravitational constant.}

\section{Summary and conclusion} \label{sec:conclusions}

The discovery of bright, compact galaxies at $z > 10$, made possible by JWST and ALMA, is fundamentally reshaping our understanding of galaxy formation at cosmic dawn. Among these sources are systems like GS-z14-0 and GHZ2, which stand out for their intense [OIII]~$88\,\mu$m emission yet weak or undetected [CII]~$158\,\mu$m lines—puzzling signatures that challenge existing theoretical models. In this work we used the high-resolution \code{SERRA} zoom-in simulations to investigate the physical origin of such extreme FIR properties and the dynamical states of early galaxies.

Our focus was Amaryllis, a synthetic analog of these [OIII]-luminous, [CII]-faint systems. Tracked from $z = 16$ to $z = 7$, Amaryllis grows from a low-mass progenitor to a $M_\star \sim 2 \times 10^{10}\,\msun$ galaxy through a sequence of hierarchical mergers and starburst episodes. By $z \sim 11.3$, it exhibits a stellar mass, SFR, compact size, and FIR/UV line luminosities that match those of GS-z14-0, making it a compelling theoretical counterpart.

We find that extreme [OIII]/[CII] ratios, reaching values as high as $\sim 8$, are not steady-state features but rather arise during brief, highly dynamic phases. These episodes are triggered by major mergers, which drive rapid gas inflows, compress the ISM, and ignite intense, bursty star formation. As a result, the ISM becomes strongly ionized ($U_{\rm ion} \sim 0.1$) and metal-poor ($Z \sim 0.02$–$0.03\,Z_\odot$), creating ideal conditions for [OIII] to dominate the FIR line emission. During these bursts, we also find evidence of powerful outflows, with $v_{\rm outflow} \sim 260\,\mathrm{kms}^{-1}$, as revealed by the broad, asymmetric wings in the integrated [OIII] spectrum and the bipolar morphology of [OIII] emission in edge-on view. The [OIII]/[CII] ratio correlates most strongly with ionization-parameter and star-formation variability, reinforcing its interpretation as a tracer of nonequilibrium conditions induced by dynamical transformations.

Yet, perhaps most surprisingly, we find that even during such a chaotic assembly, Amaryllis forms a dynamically cold gaseous disk as early as $z \sim 11$, with a $V/\sigma$ of $\sim 4$ and 6 in [OIII] and [CII], respectively. This rotationally supported gas structure persists even as the stellar component remains dispersion-dominated well into lower redshifts. This coexistence of ordered rotation with merger-driven bursts and outflows suggests a richer picture of early galaxy dynamics than often assumed.

Finally, by quantifying the DM halo of Amaryllis at $z=7.3$, we show that it resides in a massive yet not unusual $\Lambda$CDM halo, with $M_{200\mathrm{c}}\simeq1.5\times10^{11}\,\msun$ and $\lambda\simeq0.016$. The inner density profile is best described by a Burkert core with $r_c\simeq2\,\mathrm{kpc}$, corresponding to a formal concentration $c_\mathrm{Burkert}\simeq10.3$. We caution that this value is not directly comparable to the standard NFW $c$–$M$ relations; it reflects the cusp–flattening impact of baryonic feedback rather than a departure from $\Lambda$CDM. Thus, the key result is that Amaryllis lies on the high-mass tail of the expected halo distribution but is not an outlier. The combination of modest spin, early assembly, and a feedback-flattened inner profile provides a dynamically responsive potential that enables cold gas to settle into a disk long before the stellar component. This halo context reinforces the interpretation of Amaryllis as a baryon-dominated, rotation-supported system at cosmic dawn.

These findings imply that some of the brightest FIR-emitting galaxies observed at $z > 10$ have been caught in a brief, rapidly evolving phase, when mergers and feedback violently reshape the ISM and rotation begins to emerge and stabilize the gaseous component. The complexity of these systems cautions against overly simplified interpretations based on global line ratios alone and highlights the need for spatially resolved ALMA follow-up to reveal the true dynamical and morphological states of early galaxies.

In this context, Amaryllis serves as both a theoretical analog and a predictive framework, bridging the gap between compact, metal-poor galaxies observed in the FIR and the dynamically mature systems expected by the end of reionization. As future surveys push deeper into the early Universe, combining kinematic tracers, resolved line diagnostics, and multiline modeling will be essential to disentangle the interplay between mergers, outflows, and disk formation in the first galaxies.

\begin{acknowledgements}
AF acknowledges support from the ERC Advanced Grant INTERSTELLAR H2020/740120. Partial support from the Carl Friedrich von Siemens-Forschungspreis der Alexander von Humboldt-Stiftung Research Award is kindly acknowledged (AF). This research was supported (AF) in part by grant NSF PHY-2309135 to the Kavli Institute for Theoretical Physics (KITP).
We acknowledge the CINECA award under the ISCRA initiative for the availability of high-performance computing resources and support from the Class B project SERRA HP10BPUZ8F (PI: Pallottini).
We gratefully acknowledge the computational resources of the Center for High-Performance Computing (CHPC) at SNS.
We acknowledge use of the Python programming language \citep{VanRossum1991}, Astropy \citep{astropy}, Cython \citep{behnel2010cython}, Matplotlib \citep{Hunter2007}, numba \citep{numba}, NumPy \citep{VanDerWalt2011}, \code{pynbody} \citep{pynbody}, and SciPy \citep{scipy2019}.
\end{acknowledgements}

\bibliographystyle{aa_url}
\bibliography{aa55499-25}

@ARTICLE{Arata+20,
       author = {{Arata}, Shohei and {Yajima}, Hidenobu and {Nagamine}, Kentaro and {Abe}, Makito and {Khochfar}, Sadegh},
        title = "{Starbursting [O III] emitters and quiescent [C II] emitters in the reionization era}",
      journal = {\mnras},
         year = 2020,
        month = nov,
       volume = {498},
       number = {4},
        pages = {5541-5556},
          doi = {10.1093/mnras/staa2809},
archivePrefix = {arXiv},
       eprint = {2001.01853},
 primaryClass = {astro-ph.GA},
       adsurl = {https://ui.adsabs.harvard.edu/abs/2020MNRAS.498.5541A}
}

@ARTICLE{Arrabal23,
       author = {{Arrabal Haro}, Pablo and {Dickinson}, Mark and {Finkelstein}, Steven L. and {Fujimoto}, Seiji and {Fern{\'a}ndez}, Vital and {Kartaltepe}, Jeyhan S. and {Jung}, Intae and {Cole}, Justin W. and {Burgarella}, Denis and {Chworowsky}, Katherine and {Hutchison}, Taylor A. and {Morales}, Alexa M. and {Papovich}, Casey and {Simons}, Raymond C. and {Amor{\'\i}n}, Ricardo O. and {Backhaus}, Bren E. and {Bagley}, Micaela B. and {Bisigello}, Laura and {Calabr{\`o}}, Antonello and {Castellano}, Marco and {Cleri}, Nikko J. and {Dav{\'e}}, Romeel and {Dekel}, Avishai and {Ferguson}, Henry C. and {Fontana}, Adriano and {Gawiser}, Eric and {Giavalisco}, Mauro and {Harish}, Santosh and {Hathi}, Nimish P. and {Hirschmann}, Michaela and {Holwerda}, Benne W. and {Huertas-Company}, Marc and {Koekemoer}, Anton M. and {Larson}, Rebecca L. and {Lucas}, Ray A. and {Mobasher}, Bahram and {P{\'e}rez-Gonz{\'a}lez}, Pablo G. and {Pirzkal}, Nor and {Rose}, Caitlin and {Santini}, Paola and {Trump}, Jonathan R. and {de la Vega}, Alexander and {Wang}, Xin and {Weiner}, Benjamin J. and {Wilkins}, Stephen M. and {Yang}, Guang and {Yung}, L.~Y. Aaron and {Zavala}, Jorge A.},
        title = "{Spectroscopic Confirmation of CEERS NIRCam-selected Galaxies at z ≃ 8-10}",
      journal = {\apjl},
         year = 2023,
        month = jul,
       volume = {951},
       number = {1},
          eid = {L22},
        pages = {L22},
          doi = {10.3847/2041-8213/acdd54},
archivePrefix = {arXiv},
       eprint = {2304.05378},
 primaryClass = {astro-ph.GA},
       adsurl = {https://ui.adsabs.harvard.edu/abs/2023ApJ...951L..22A}
}

@ARTICLE{astropy,
   author = {{Astropy Collaboration} and {Robitaille}, T.~P. and {Tollerud}, E.~J. and 
	{Greenfield}, P. and {Droettboom}, M. and {Bray}, E. and {Aldcroft}, T. and 
	{Davis}, M. and {Ginsburg}, A. and {Price-Whelan}, A.~M. and 
	{Kerzendorf}, W.~E. and {Conley}, A. and {Crighton}, N. and 
	{Barbary}, K. and {Muna}, D. and {Ferguson}, H. and {Grollier}, F. and 
	{Parikh}, M.~M. and {Nair}, P.~H. and {Unther}, H.~M. and {Deil}, C. and 
	{Woillez}, J. and {Conseil}, S. and {Kramer}, R. and {Turner}, J.~E.~H. and 
	{Singer}, L. and {Fox}, R. and {Weaver}, B.~A. and {Zabalza}, V. and 
	{Edwards}, Z.~I. and {Azalee Bostroem}, K. and {Burke}, D.~J. and 
	{Casey}, A.~R. and {Crawford}, S.~M. and {Dencheva}, N. and 
	{Ely}, J. and {Jenness}, T. and {Labrie}, K. and {Lim}, P.~L. and 
	{Pierfederici}, F. and {Pontzen}, A. and {Ptak}, A. and {Refsdal}, B. and 
	{Servillat}, M. and {Streicher}, O.},
    title = "{Astropy: A community Python package for astronomy}",
  journal = {\aap},
archivePrefix = "arXiv",
   eprint = {1307.6212},
 primaryClass = "astro-ph.IM",
     year = 2013,
   volume = 558,
      eid = {A33},
    pages = {A33},
      doi = {10.1051/0004-6361/201322068},
   adsurl = {http://adsabs.harvard.edu/abs/2013A%26A...558A..33A}
}

@ARTICLE{atek:2023,
       author = {{Atek}, Hakim and {Shuntov}, Marko and {Furtak}, Lukas J. and {Richard}, Johan and {Kneib}, Jean-Paul and {Mahler}, Guillaume and {Zitrin}, Adi and {McCracken}, H.~J. and {Charlot}, St{\'e}phane and {Chevallard}, Jacopo and {Chemerynska}, Iryna},
        title = "{Revealing galaxy candidates out to z   16 with JWST observations of the lensing cluster SMACS0723}",
      journal = {\mnras},
         year = 2023,
        month = feb,
       volume = {519},
       number = {1},
        pages = {1201-1220},
          doi = {10.1093/mnras/stac3144},
archivePrefix = {arXiv},
       eprint = {2207.12338},
 primaryClass = {astro-ph.GA},
       adsurl = {https://ui.adsabs.harvard.edu/abs/2023MNRAS.519.1201A}
}

@ARTICLE{Baes2015,
   author = {{Baes}, M. and {Camps}, P.},
    title = "{SKIRT: The design of a suite of input models for Monte Carlo radiative transfer simulations}",
  journal = {Astronomy and Computing},
archivePrefix = "arXiv",
   eprint = {1505.07708},
 primaryClass = "astro-ph.IM",
     year = 2015,
    month = sep,
   volume = 12,
    pages = {33-44},
      doi = {10.1016/j.ascom.2015.05.006},
   adsurl = {http://adsabs.harvard.edu/abs/2015A%26C....12...33B}
}

@ARTICLE{Bakx22,
       author = {{Bakx}, Tom J.~L.~C. and {Zavala}, Jorge A. and {Mitsuhashi}, Ikki and {Treu}, Tommaso and {Fontana}, Adriano and {Tadaki}, Ken-ichi and {Casey}, Caitlin M. and {Castellano}, Marco and {Glazebrook}, Karl and {Hagimoto}, Masato and {Ikeda}, Ryota and {Jones}, Tucker and {Leethochawalit}, Nicha and {Mason}, Charlotte and {Morishita}, Takahiro and {Nanayakkara}, Themiya and {Pentericci}, Laura and {Roberts-Borsani}, Guido and {Santini}, Paola and {Serjeant}, Stephen and {Tamura}, Yoichi and {Trenti}, Michele and {Vanzella}, Eros},
        title = "{Deep ALMA redshift search of a z {\ensuremath{\sim}} 12 GLASS-JWST galaxy candidate}",
      journal = {\mnras},
         year = 2023,
        month = mar,
       volume = {519},
       number = {4},
        pages = {5076-5085},
          doi = {10.1093/mnras/stac3723},
archivePrefix = {arXiv},
       eprint = {2208.13642},
 primaryClass = {astro-ph.GA},
       adsurl = {https://ui.adsabs.harvard.edu/abs/2023MNRAS.519.5076B}
}

@ARTICLE{Barnes+86,
       author = {{Barnes}, Josh and {Hut}, Piet},
        title = "{A hierarchical O(N log N) force-calculation algorithm}",
      journal = {\nat},
         year = 1986,
        month = dec,
       volume = {324},
       number = {6096},
        pages = {446-449},
          doi = {10.1038/324446a0},
       adsurl = {https://ui.adsabs.harvard.edu/abs/1986Natur.324..446B}
}

@ARTICLE{behnel2010cython,
    author={Behnel, S. and Bradshaw, R. and Citro, C. and Dalcin, L. and Seljebotn, D.S. and Smith, K.},
    journal={Computing in Science Engineering},
    title={Cython: The Best of Both Worlds},
    year={2011},
    month=march-april ,
    volume={13},
    number={2},
    pages={31 -39},
    doi={10.1109/MCSE.2010.118},
    ISSN={1521-9615},
}

@article{behrens:2018dust,
 adsurl = {https://ui.adsabs.harvard.edu/#abs/2018MNRAS.477..552B},
 author = {{Behrens}, C. and {Pallottini}, A. and {Ferrara}, A. and {Gallerani}, S.
and {Vallini}, L.},
 doi = {10.1093/mnras/sty552},
 journal = {\mnras},
 month = {June},
 pages = {552-565},
 title = {{Dusty galaxies in the Epoch of Reionization: simulations}},
 volume = {477},
 year = {2018}
}

@Article{behroozi:2013,
  author    = {{Behroozi}, Peter S. and {Wechsler}, Risa H. and {Wu},
          Hao-Yi},
  title     = "{The ROCKSTAR Phase-space Temporal Halo Finder and the
          Velocity Offsets of Cluster Cores}",
  journal   = {\apj},
  year      = 2013,
  month     = jan,
  volume    = {762},
  number    = {2},
  eid       = {109},
  pages     = {109},
  doi       = {10.1088/0004-637X/762/2/109},
  archiveprefix = {arXiv},
  eprint    = {1110.4372},
  primaryclass  = {astro-ph.CO},
  adsurl    = {https://ui.adsabs.harvard.edu/abs/2013ApJ...762..109B}
}

@ARTICLE{Bullock2001,
       author = {{Bullock}, J.~S. and {Dekel}, A. and {Kolatt}, T.~S. and {Kravtsov}, A.~V. and {Klypin}, A.~A. and {Porciani}, C. and {Primack}, J.~R.},
        title = "{A Universal Angular Momentum Profile for Galactic Halos}",
      journal = {\apj},
         year = 2001,
        month = jul,
       volume = {555},
       number = {1},
        pages = {240-257},
          doi = {10.1086/321477},
archivePrefix = {arXiv},
       eprint = {astro-ph/0011001},
 primaryClass = {astro-ph},
       adsurl = {https://ui.adsabs.harvard.edu/abs/2001ApJ...555..240B}
}

@ARTICLE{Bunker23,
       author = {{Bunker}, Andrew J. and {Saxena}, Aayush and {Cameron}, Alex J. and {Willott}, Chris J. and {Curtis-Lake}, Emma and {Jakobsen}, Peter and {Carniani}, Stefano and {Smit}, Renske and {Maiolino}, Roberto and {Witstok}, Joris and {Curti}, Mirko and {D'Eugenio}, Francesco and {Jones}, Gareth C. and {Ferruit}, Pierre and {Arribas}, Santiago and {Charlot}, Stephane and {Chevallard}, Jacopo and {Giardino}, Giovanna and {de Graaff}, Anna and {Looser}, Tobias J. and {L{\"u}tzgendorf}, Nora and {Maseda}, Michael V. and {Rawle}, Tim and {Rix}, Hans-Walter and {Del Pino}, Bruno Rodr{\'\i}guez and {Alberts}, Stacey and {Egami}, Eiichi and {Eisenstein}, Daniel J. and {Endsley}, Ryan and {Hainline}, Kevin and {Hausen}, Ryan and {Johnson}, Benjamin D. and {Rieke}, George and {Rieke}, Marcia and {Robertson}, Brant E. and {Shivaei}, Irene and {Stark}, Daniel P. and {Sun}, Fengwu and {Tacchella}, Sandro and {Tang}, Mengtao and {Williams}, Christina C. and {Willmer}, Christopher N.~A. and {Baker}, William M. and {Baum}, Stefi and {Bhatawdekar}, Rachana and {Bowler}, Rebecca and {Boyett}, Kristan and {Chen}, Zuyi and {Circosta}, Chiara and {Helton}, Jakob M. and {Ji}, Zhiyuan and {Kumari}, Nimisha and {Lyu}, Jianwei and {Nelson}, Erica and {Parlanti}, Eleonora and {Perna}, Michele and {Sandles}, Lester and {Scholtz}, Jan and {Suess}, Katherine A. and {Topping}, Michael W. and {{\"U}bler}, Hannah and {Wallace}, Imaan E.~B. and {Whitler}, Lily},
        title = "{JADES NIRSpec Spectroscopy of GN-z11: Lyman-{\ensuremath{\alpha}} emission and possible enhanced nitrogen abundance in a z = 10.60 luminous galaxy}",
      journal = {\aap},
         year = 2023,
        month = sep,
       volume = {677},
          eid = {A88},
        pages = {A88},
          doi = {10.1051/0004-6361/202346159},
archivePrefix = {arXiv},
       eprint = {2302.07256},
 primaryClass = {astro-ph.GA},
       adsurl = {https://ui.adsabs.harvard.edu/abs/2023A&A...677A..88B}
}

@ARTICLE{Buowens+21,
       author = {{Bouwens}, R.~J. and {Smit}, R. and {Schouws}, S. and {Stefanon}, M. and {Bowler}, R. and {Endsley}, R. and {Gonzalez}, V. and {Inami}, H. and {Stark}, D. and {Oesch}, P. and {Hodge}, J. and {Aravena}, M. and {da Cunha}, E. and {Dayal}, P. and {de Looze}, I. and {Ferrara}, A. and {Fudamoto}, Y. and {Graziani}, L. and {Li}, C. and {Nanayakkara}, T. and {Pallottini}, A. and {Schneider}, R. and {Sommovigo}, L. and {Topping}, M. and {van der Werf}, P. and {Algera}, H. and {Barrufet}, L. and {Hygate}, A. and {Labb{\'e}}, I. and {Riechers}, D. and {Witstok}, J.},
        title = "{Reionization Era Bright Emission Line Survey: Selection and Characterization of Luminous Interstellar Medium Reservoirs in the z > 6.5 Universe}",
      journal = {\apj},
         year = 2022,
        month = jun,
       volume = {931},
       number = {2},
          eid = {160},
        pages = {160},
          doi = {10.3847/1538-4357/ac5a4a},
archivePrefix = {arXiv},
       eprint = {2106.13719},
 primaryClass = {astro-ph.GA},
       adsurl = {https://ui.adsabs.harvard.edu/abs/2022ApJ...931..160B}
}

@ARTICLE{Burkert1995,
       author = {{Burkert}, A.},
        title = "{The Structure of Dark Matter Halos in Dwarf Galaxies}",
      journal = {\apjl},
         year = 1995,
        month = jul,
       volume = {447},
        pages = {L25-L28},
          doi = {10.1086/309560},
archivePrefix = {arXiv},
       eprint = {astro-ph/9504041},
 primaryClass = {astro-ph},
       adsurl = {https://ui.adsabs.harvard.edu/abs/1995ApJ...447L..25B}
}

@ARTICLE{Calabro+24,
       author = {{Calabr{\`o}}, Antonello and {Castellano}, Marco and {Zavala}, Jorge A. and {Pentericci}, Laura and {Arrabal Haro}, Pablo and {Bakx}, Tom J.~L.~C. and {Burgarella}, Denis and {Casey}, Caitlin M. and {Dickinson}, Mark and {Finkelstein}, Steven L. and {Fontana}, Adriano and {Llerena}, Mario and {Mascia}, Sara and {Merlin}, Emiliano and {Mitsuhashi}, Ikki and {Napolitano}, Lorenzo and {Paris}, Diego and {P{\'e}rez-Gonz{\'a}lez}, Pablo G. and {Roberts-Borsani}, Guido and {Santini}, Paola and {Treu}, Tommaso and {Vanzella}, Eros},
        title = "{Evidence of Extreme Ionization Conditions and Low Metallicity in GHZ2/GLASS-Z12 from a Combined Analysis of NIRSpec and MIRI Observations}",
      journal = {\apj},
         year = 2024,
        month = nov,
       volume = {975},
       number = {2},
          eid = {245},
        pages = {245},
          doi = {10.3847/1538-4357/ad7602},
archivePrefix = {arXiv},
       eprint = {2403.12683},
 primaryClass = {astro-ph.GA},
       adsurl = {https://ui.adsabs.harvard.edu/abs/2024ApJ...975..245C}
}

@ARTICLE{Camps2015,
   author = {{Camps}, P. and {Baes}, M.},
    title = "{SKIRT: An advanced dust radiative transfer code with a user-friendly architecture}",
  journal = {Astronomy and Computing},
archivePrefix = "arXiv",
   eprint = {1410.1629},
 primaryClass = "astro-ph.IM",
     year = 2015,
    month = mar,
   volume = 9,
    pages = {20-33},
      doi = {10.1016/j.ascom.2014.10.004},
   adsurl = {http://adsabs.harvard.edu/abs/2015A%26C.....9...20C}
}

@ARTICLE{Carniani+20,
       author = {{Carniani}, S. and {Ferrara}, A. and {Maiolino}, R. and {Castellano}, M. and {Gallerani}, S. and {Fontana}, A. and {Kohandel}, M. and {Lupi}, A. and {Pallottini}, A. and {Pentericci}, L. and {Vallini}, L. and {Vanzella}, E.},
        title = "{Missing [C II] emission from early galaxies}",
      journal = {\mnras},
         year = 2020,
        month = dec,
       volume = {499},
       number = {4},
        pages = {5136-5150},
          doi = {10.1093/mnras/staa3178},
archivePrefix = {arXiv},
       eprint = {2006.09402},
 primaryClass = {astro-ph.GA},
       adsurl = {https://ui.adsabs.harvard.edu/abs/2020MNRAS.499.5136C}
}

@ARTICLE{Carniani+24,
       author = {{Carniani}, Stefano and {Venturi}, Giacomo and {Parlanti}, Eleonora and {de Graaff}, Anna and {Maiolino}, Roberto and {Arribas}, Santiago and {Bonaventura}, Nina and {Boyett}, Kristan and {Bunker}, Andrew J. and {Cameron}, Alex J. and {Charlot}, Stephane and {Chevallard}, Jacopo and {Curti}, Mirko and {Curtis-Lake}, Emma and {Eisenstein}, Daniel J. and {Giardino}, Giovanna and {Hausen}, Ryan and {Kumari}, Nimisha and {Maseda}, Michael V. and {Nelson}, Erica and {Perna}, Michele and {Rix}, Hans-Walter and {Robertson}, Brant and {Del Pino}, Bruno Rodr{\'\i}guez and {Sandles}, Lester and {Scholtz}, Jan and {Simmonds}, Charlotte and {Smit}, Renske and {Tacchella}, Sandro and {{\"U}bler}, Hannah and {Williams}, Christina C. and {Willott}, Chris and {Witstok}, Joris},
        title = "{JADES: The incidence rate and properties of galactic outflows in low-mass galaxies across 3 < z < 9}",
      journal = {\aap},
         year = 2024,
        month = may,
       volume = {685},
          eid = {A99},
        pages = {A99},
          doi = {10.1051/0004-6361/202347230},
archivePrefix = {arXiv},
       eprint = {2306.11801},
 primaryClass = {astro-ph.GA},
       adsurl = {https://ui.adsabs.harvard.edu/abs/2024A&A...685A..99C}
}

@ARTICLE{Carniani+24z14alma,
       author = {{Carniani}, Stefano and {D'Eugenio}, Francesco and {Ji}, Xihan and {Parlanti}, Eleonora and {Scholtz}, Jan and {Sun}, Fengwu and {Venturi}, Giacomo and {Bakx}, Tom J.~L.~C. and {Curti}, Mirko and {Maiolino}, Roberto and {Tacchella}, Sandro and {Zavala}, Jorge A. and {Hainline}, Kevin and {Witstok}, Joris and {Johnson}, Benjamin D. and {Alberts}, Stacey and {Bunker}, Andrew J. and {Charlot}, St{\'e}phane and {Eisenstein}, Daniel J. and {Helton}, Jakob M. and {Jakobsen}, Peter and {Kumari}, Nimisha and {Robertson}, Brant and {Saxena}, Aayush and {{\"U}bler}, Hannah and {Williams}, Christina C. and {Willmer}, Christopher N.~A. and {Willott}, Chris},
        title = "{The eventful life of a luminous galaxy at z = 14: metal enrichment, feedback, and low gas fraction?}",
      journal = {\aap},
         year = 2025,
        month = apr,
       volume = {696},
          eid = {A87},
        pages = {A87},
          doi = {10.1051/0004-6361/202452451},
archivePrefix = {arXiv},
       eprint = {2409.20533},
 primaryClass = {astro-ph.GA},
       adsurl = {https://ui.adsabs.harvard.edu/abs/2025A&A...696A..87C}
}

@ARTICLE{Carniani+24z14nirspec,
       author = {{Carniani}, Stefano and {Hainline}, Kevin and {D'Eugenio}, Francesco and {Eisenstein}, Daniel J. and {Jakobsen}, Peter and {Witstok}, Joris and {Johnson}, Benjamin D. and {Chevallard}, Jacopo and {Maiolino}, Roberto and {Helton}, Jakob M. and {Willott}, Chris and {Robertson}, Brant and {Alberts}, Stacey and {Arribas}, Santiago and {Baker}, William M. and {Bhatawdekar}, Rachana and {Boyett}, Kristan and {Bunker}, Andrew J. and {Cameron}, Alex J. and {Cargile}, Phillip A. and {Charlot}, St{\'e}phane and {Curti}, Mirko and {Curtis-Lake}, Emma and {Egami}, Eiichi and {Giardino}, Giovanna and {Isaak}, Kate and {Ji}, Zhiyuan and {Jones}, Gareth C. and {Kumari}, Nimisha and {Maseda}, Michael V. and {Parlanti}, Eleonora and {P{\'e}rez-Gonz{\'a}lez}, Pablo G. and {Rawle}, Tim and {Rieke}, George and {Rieke}, Marcia and {Del Pino}, Bruno Rodr{\'\i}guez and {Saxena}, Aayush and {Scholtz}, Jan and {Smit}, Renske and {Sun}, Fengwu and {Tacchella}, Sandro and {{\"U}bler}, Hannah and {Venturi}, Giacomo and {Williams}, Christina C. and {Willmer}, Christopher N.~A.},
        title = "{Spectroscopic confirmation of two luminous galaxies at a redshift of 14}",
      journal = {\nat},
         year = 2024,
        month = sep,
       volume = {633},
       number = {8029},
        pages = {318-322},
          doi = {10.1038/s41586-024-07860-9},
archivePrefix = {arXiv},
       eprint = {2405.18485},
 primaryClass = {astro-ph.GA},
       adsurl = {https://ui.adsabs.harvard.edu/abs/2024Natur.633..318C}
}

@ARTICLE{Castellano+24,
       author = {{Castellano}, Marco and {Napolitano}, Lorenzo and {Fontana}, Adriano and {Roberts-Borsani}, Guido and {Treu}, Tommaso and {Vanzella}, Eros and {Zavala}, Jorge A. and {Arrabal Haro}, Pablo and {Calabr{\`o}}, Antonello and {Llerena}, Mario and {Mascia}, Sara and {Merlin}, Emiliano and {Paris}, Diego and {Pentericci}, Laura and {Santini}, Paola and {Bakx}, Tom J.~L.~C. and {Bergamini}, Pietro and {Cupani}, Guido and {Dickinson}, Mark and {Filippenko}, Alexei V. and {Glazebrook}, Karl and {Grillo}, Claudio and {Kelly}, Patrick L. and {Malkan}, Matthew A. and {Mason}, Charlotte A. and {Morishita}, Takahiro and {Nanayakkara}, Themiya and {Rosati}, Piero and {Sani}, Eleonora and {Wang}, Xin and {Yoon}, Ilsang},
        title = "{JWST NIRSpec Spectroscopy of the Remarkable Bright Galaxy GHZ2/GLASS-z12 at Redshift 12.34}",
      journal = {\apj},
         year = 2024,
        month = sep,
       volume = {972},
       number = {2},
          eid = {143},
        pages = {143},
          doi = {10.3847/1538-4357/ad5f88},
archivePrefix = {arXiv},
       eprint = {2403.10238},
 primaryClass = {astro-ph.GA},
       adsurl = {https://ui.adsabs.harvard.edu/abs/2024ApJ...972..143C}
}

@ARTICLE{Castellano22,
       author = {{Castellano}, Marco and {Fontana}, Adriano and {Treu}, Tommaso and {Santini}, Paola and {Merlin}, Emiliano and {Leethochawalit}, Nicha and {Trenti}, Michele and {Vanzella}, Eros and {Mestric}, Uros and {Bonchi}, Andrea and {Belfiori}, Davide and {Nonino}, Mario and {Paris}, Diego and {Polenta}, Gianluca and {Roberts-Borsani}, Guido and {Boyett}, Kristan and {Brada{\v{c}}}, Maru{\v{s}}a and {Calabr{\`o}}, Antonello and {Glazebrook}, Karl and {Grillo}, Claudio and {Mascia}, Sara and {Mason}, Charlotte and {Mercurio}, Amata and {Morishita}, Takahiro and {Nanayakkara}, Themiya and {Pentericci}, Laura and {Rosati}, Piero and {Vulcani}, Benedetta and {Wang}, Xin and {Yang}, Lilan},
        title = "{Early Results from GLASS-JWST. III. Galaxy Candidates at z  9-15}",
      journal = {\apjl},
         year = 2022,
        month = oct,
       volume = {938},
       number = {2},
          eid = {L15},
        pages = {L15},
          doi = {10.3847/2041-8213/ac94d0},
archivePrefix = {arXiv},
       eprint = {2207.09436},
 primaryClass = {astro-ph.GA},
       adsurl = {https://ui.adsabs.harvard.edu/abs/2022ApJ...938L..15C}
}

@ARTICLE{Castellano24,
       author = {{Castellano}, Marco and {Napolitano}, Lorenzo and {Fontana}, Adriano and {Roberts-Borsani}, Guido and {Treu}, Tommaso and {Vanzella}, Eros and {Zavala}, Jorge A. and {Arrabal Haro}, Pablo and {Calabr{\`o}}, Antonello and {Llerena}, Mario and {Mascia}, Sara and {Merlin}, Emiliano and {Paris}, Diego and {Pentericci}, Laura and {Santini}, Paola and {Bakx}, Tom J.~L.~C. and {Bergamini}, Pietro and {Cupani}, Guido and {Dickinson}, Mark and {Filippenko}, Alexei V. and {Glazebrook}, Karl and {Grillo}, Claudio and {Kelly}, Patrick L. and {Malkan}, Matthew A. and {Mason}, Charlotte A. and {Morishita}, Takahiro and {Nanayakkara}, Themiya and {Rosati}, Piero and {Sani}, Eleonora and {Wang}, Xin and {Yoon}, Ilsang},
        title = "{JWST NIRSpec Spectroscopy of the Remarkable Bright Galaxy GHZ2/GLASS-z12 at Redshift 12.34}",
      journal = {\apj},
         year = 2024,
        month = sep,
       volume = {972},
       number = {2},
          eid = {143},
        pages = {143},
          doi = {10.3847/1538-4357/ad5f88},
archivePrefix = {arXiv},
       eprint = {2403.10238},
 primaryClass = {astro-ph.GA},
       adsurl = {https://ui.adsabs.harvard.edu/abs/2024ApJ...972..143C}
}

@ARTICLE{Ceverino+15,
       author = {{Ceverino}, Daniel and {Dekel}, Avishai and {Tweed}, Dylan and {Primack}, Joel},
        title = "{Early formation of massive, compact, spheroidal galaxies with classical profiles by violent disc instability or mergers}",
      journal = {\mnras},
         year = 2015,
        month = mar,
       volume = {447},
       number = {4},
        pages = {3291-3310},
          doi = {10.1093/mnras/stu2694},
archivePrefix = {arXiv},
       eprint = {1409.2622},
 primaryClass = {astro-ph.GA},
       adsurl = {https://ui.adsabs.harvard.edu/abs/2015MNRAS.447.3291C}
}

@ARTICLE{Curtis-Lake+23,
       author = {{Curtis-Lake}, Emma and {Carniani}, Stefano and {Cameron}, Alex and {Charlot}, Stephane and {Jakobsen}, Peter and {Maiolino}, Roberto and {Bunker}, Andrew and {Witstok}, Joris and {Smit}, Renske and {Chevallard}, Jacopo and {Willott}, Chris and {Ferruit}, Pierre and {Arribas}, Santiago and {Bonaventura}, Nina and {Curti}, Mirko and {D'Eugenio}, Francesco and {Franx}, Marijn and {Giardino}, Giovanna and {Looser}, Tobias J. and {L{\"u}tzgendorf}, Nora and {Maseda}, Michael V. and {Rawle}, Tim and {Rix}, Hans-Walter and {Rodr{\'\i}guez del Pino}, Bruno and {{\"U}bler}, Hannah and {Sirianni}, Marco and {Dressler}, Alan and {Egami}, Eiichi and {Eisenstein}, Daniel J. and {Endsley}, Ryan and {Hainline}, Kevin and {Hausen}, Ryan and {Johnson}, Benjamin D. and {Rieke}, Marcia and {Robertson}, Brant and {Shivaei}, Irene and {Stark}, Daniel P. and {Tacchella}, Sandro and {Williams}, Christina C. and {Willmer}, Christopher N.~A. and {Bhatawdekar}, Rachana and {Bowler}, Rebecca and {Boyett}, Kristan and {Chen}, Zuyi and {de Graaff}, Anna and {Helton}, Jakob M. and {Hviding}, Raphael E. and {Jones}, Gareth C. and {Kumari}, Nimisha and {Lyu}, Jianwei and {Nelson}, Erica and {Perna}, Michele and {Sandles}, Lester and {Saxena}, Aayush and {Suess}, Katherine A. and {Sun}, Fengwu and {Topping}, Michael W. and {Wallace}, Imaan E.~B. and {Whitler}, Lily},
        title = "{Spectroscopic confirmation of four metal-poor galaxies at z = 10.3-13.2}",
      journal = {Nature Astronomy},
         year = 2023,
        month = may,
       volume = {7},
        pages = {622-632},
          doi = {10.1038/s41550-023-01918-w},
archivePrefix = {arXiv},
       eprint = {2212.04568},
 primaryClass = {astro-ph.GA},
       adsurl = {https://ui.adsabs.harvard.edu/abs/2023NatAs...7..622C}
}

@ARTICLE{Dayal+22,
       author = {{Dayal}, P. and {Ferrara}, A. and {Sommovigo}, L. and {Bouwens}, R. and {Oesch}, P.~A. and {Smit}, R. and {Gonzalez}, V. and {Schouws}, S. and {Stefanon}, M. and {Kobayashi}, C. and {Bremer}, J. and {Algera}, H.~S.~B. and {Aravena}, M. and {Bowler}, R.~A.~A. and {da Cunha}, E. and {Fudamoto}, Y. and {Graziani}, L. and {Hodge}, J. and {Inami}, H. and {De Looze}, I. and {Pallottini}, A. and {Riechers}, D. and {Schneider}, R. and {Stark}, D. and {Endsley}, R.},
        title = "{The ALMA REBELS survey: the dust content of z   7 Lyman break galaxies}",
      journal = {\mnras},
         year = 2022,
        month = may,
       volume = {512},
       number = {1},
        pages = {989-1002},
          doi = {10.1093/mnras/stac537},
archivePrefix = {arXiv},
       eprint = {2202.11118},
 primaryClass = {astro-ph.GA},
       adsurl = {https://ui.adsabs.harvard.edu/abs/2022MNRAS.512..989D}
}

@article{decataldo:2019,
       author = {{Decataldo}, D. and {Pallottini}, A. and {Ferrara}, A. and
         {Vallini}, L. and {Gallerani}, S.},
        title = "{Photoevaporation of Jeans-unstable molecular clumps}",
      journal = {\mnras},
         year = "2019",
        month = "Aug",
       volume = {487},
       number = {3},
        pages = {3377-3391},
          doi = {10.1093/mnras/stz1527},
archivePrefix = {arXiv},
       eprint = {1905.13230},
 primaryClass = {astro-ph.GA},
       adsurl = {https://ui.adsabs.harvard.edu/abs/2019MNRAS.487.3377D}
}

@ARTICLE{DeLooze+14,
   author = {{De Looze}, I. and {Cormier}, D. and {Lebouteiller}, V. and 
	{Madden}, S. and {Baes}, M. and {Bendo}, G.~J. and {Boquien}, M. and 
	{Boselli}, A. and {Clements}, D.~L. and {Cortese}, L. and {Cooray}, A. and 
	{Galametz}, M. and {Galliano}, F. and {Graci{\'a}-Carpio}, J. and 
	{Isaak}, K. and {Karczewski}, O.~{\L}. and {Parkin}, T.~J. and 
	{Pellegrini}, E.~W. and {R{\'e}my-Ruyer}, A. and {Spinoglio}, L. and 
	{Smith}, M.~W.~L. and {Sturm}, E.},
    title = "{The applicability of far-infrared fine-structure lines as star formation rate tracers over wide ranges of metallicities and galaxy types}",
  journal = {\aap},
archivePrefix = "arXiv",
   eprint = {1402.4075},
     year = 2014,
    month = aug,
   volume = 568,
      eid = {A62},
    pages = {A62},
      doi = {10.1051/0004-6361/201322489},
   adsurl = {http://adsabs.harvard.edu/abs/2014A%26A...568A..62D}
}

@ARTICLE{dimascia:2025,
       author = {{Di Mascia}, Fabio and {Pallottini}, Andrea and {Sommovigo}, Laura and {Decataldo}, Davide},
        title = "{Investigating ultraviolet and infrared radiation through the turbulent life of molecular clouds}",
      journal = {\aap},
         year = 2025,
        month = mar,
       volume = {695},
          eid = {A77},
        pages = {A77},
          doi = {10.1051/0004-6361/202451430},
archivePrefix = {arXiv},
       eprint = {2407.01662},
 primaryClass = {astro-ph.GA},
       adsurl = {https://ui.adsabs.harvard.edu/abs/2025A&A...695A..77D}
}

@misc{Donnan22,
  doi = {10.48550/ARXIV.2207.12356},
  
  url = {https://arxiv.org/abs/2207.12356},
  
  author = {Donnan, C. T. and McLeod, D. J. and Dunlop, J. S. and McLure, R. J. and Carnall, A. C. and Begley, R. and Cullen, F. and Hamadouche, M. L. and Bowler, R. A. A. and McCracken, H. J. and Milvang-Jensen, B. and Moneti, A. and Targett, T.},
  
  title = {The evolution of the galaxy UV luminosity function at redshifts z ~ 8-15 from deep JWST and ground-based near-infrared imaging},
  
  publisher = {arXiv},
  
  year = {2022},
  
  copyright = {Creative Commons Attribution 4.0 International}
}

@ARTICLE{Dutton2014,
       author = {{Dutton}, Aaron A. and {Macci{\`o}}, Andrea V.},
        title = "{Cold dark matter haloes in the Planck era: evolution of structural parameters for Einasto and NFW profiles}",
      journal = {\mnras},
         year = 2014,
        month = jul,
       volume = {441},
       number = {4},
        pages = {3359-3374},
          doi = {10.1093/mnras/stu742},
archivePrefix = {arXiv},
       eprint = {1402.7073},
 primaryClass = {astro-ph.CO},
       adsurl = {https://ui.adsabs.harvard.edu/abs/2014MNRAS.441.3359D}
}

@ARTICLE{endsley:2024,
       author = {{Endsley}, Ryan and {Chisholm}, John and {Stark}, Daniel P. and {Topping}, Michael W. and {Whitler}, Lily},
        title = "{The Burstiness of Star Formation at z {\ensuremath{\sim}} 6: A Huge Diversity in the Recent Star Formation Histories of Very UV-faint Galaxies}",
      journal = {\apj},
         year = 2025,
        month = jul,
       volume = {987},
       number = {2},
          eid = {189},
        pages = {189},
          doi = {10.3847/1538-4357/addc74},
archivePrefix = {arXiv},
       eprint = {2410.01905},
 primaryClass = {astro-ph.GA},
       adsurl = {https://ui.adsabs.harvard.edu/abs/2025ApJ...987..189E}
}

@ARTICLE{Ferland+17,
       author = {{Ferland}, G.~J. and {Chatzikos}, M. and {Guzm{\'a}n}, F. and {Lykins}, M.~L. and {van Hoof}, P.~A.~M. and {Williams}, R.~J.~R. and {Abel}, N.~P. and {Badnell}, N.~R. and {Keenan}, F.~P. and {Porter}, R.~L. and {Stancil}, P.~C.},
        title = "{The 2017 Release Cloudy}",
      journal = {\rmxaa},
         year = 2017,
        month = oct,
       volume = {53},
        pages = {385-438},
          doi = {10.48550/arXiv.1705.10877},
archivePrefix = {arXiv},
       eprint = {1705.10877},
 primaryClass = {astro-ph.GA},
       adsurl = {https://ui.adsabs.harvard.edu/abs/2017RMxAA..53..385F}
}

@ARTICLE{Ferrara+23,
       author = {{Ferrara}, Andrea and {Pallottini}, Andrea and {Dayal}, Pratika},
        title = "{On the stunning abundance of super-early, luminous galaxies revealed by JWST}",
      journal = {\mnras},
         year = 2023,
        month = jul,
       volume = {522},
       number = {3},
        pages = {3986-3991},
          doi = {10.1093/mnras/stad1095},
archivePrefix = {arXiv},
       eprint = {2208.00720},
 primaryClass = {astro-ph.GA},
       adsurl = {https://ui.adsabs.harvard.edu/abs/2023MNRAS.522.3986F}
}

@ARTICLE{Ferrara+24,
       author = {{Ferrara}, A. and {Carniani}, S. and {di Mascia}, F. and {Bouwens}, R.~J. and {Oesch}, P. and {Schouws}, S.},
        title = "{ALMA observations of super-early galaxies: Attenuation-free model predictions}",
      journal = {\aap},
         year = 2025,
        month = feb,
       volume = {694},
          eid = {A215},
        pages = {A215},
          doi = {10.1051/0004-6361/202452368},
archivePrefix = {arXiv},
       eprint = {2409.17223},
 primaryClass = {astro-ph.GA},
       adsurl = {https://ui.adsabs.harvard.edu/abs/2025A&A...694A.215F}
}

@ARTICLE{Ferrara24a,
       author = {{Ferrara}, A.},
        title = "{Super-early JWST galaxies, outflows, and Ly{\ensuremath{\alpha}} visibility in the Epoch of Reionization}",
      journal = {\aap},
         year = 2024,
        month = apr,
       volume = {684},
          eid = {A207},
        pages = {A207},
          doi = {10.1051/0004-6361/202348321},
archivePrefix = {arXiv},
       eprint = {2310.12197},
 primaryClass = {astro-ph.GA},
       adsurl = {https://ui.adsabs.harvard.edu/abs/2024A&A...684A.207F}
}

@article{Ferrara:2019,
       author = {{Ferrara}, A. and {Vallini}, L. and {Pallottini}, A. and
         {Gallerani}, S. and {Carniani}, S. and {Kohandel}, M. and
         {Decataldo}, D. and {Behrens}, C.},
        title = "{A physical model for [C II] line emission from galaxies}",
      journal = {\mnras},
         year = "2019",
        month = "Oct",
       volume = {489},
       number = {1},
        pages = {1-12},
          doi = {10.1093/mnras/stz2031},
archivePrefix = {arXiv},
       eprint = {1908.07536},
 primaryClass = {astro-ph.GA},
       adsurl = {https://ui.adsabs.harvard.edu/abs/2019MNRAS.489....1F}
}

@ARTICLE{Ferreira+22a,
       author = {{Ferreira}, Leonardo and {Adams}, Nathan and {Conselice}, Christopher J. and {Sazonova}, Elizaveta and {Austin}, Duncan and {Caruana}, Joseph and {Ferrari}, Fabricio and {Verma}, Aprajita and {Trussler}, James and {Broadhurst}, Tom and {Diego}, Jose and {Frye}, Brenda L. and {Pascale}, Massimo and {Wilkins}, Stephen M. and {Windhorst}, Rogier A. and {Zitrin}, Adi},
        title = "{Panic! at the Disks: First Rest-frame Optical Observations of Galaxy Structure at z > 3 with JWST in the SMACS 0723 Field}",
      journal = {\apjl},
         year = 2022,
        month = oct,
       volume = {938},
       number = {1},
          eid = {L2},
        pages = {L2},
          doi = {10.3847/2041-8213/ac947c},
archivePrefix = {arXiv},
       eprint = {2207.09428},
 primaryClass = {astro-ph.GA},
       adsurl = {https://ui.adsabs.harvard.edu/abs/2022ApJ...938L...2F}
}

@ARTICLE{Finkelstein23,
       author = {{Finkelstein}, Steven L. and {Leung}, Gene C.~K. and {Bagley}, Micaela B. and {Dickinson}, Mark and {Ferguson}, Henry C. and {Papovich}, Casey and {Akins}, Hollis B. and {Arrabal Haro}, Pablo and {Dav{\'e}}, Romeel and {Dekel}, Avishai and {Kartaltepe}, Jeyhan S. and {Kocevski}, Dale D. and {Koekemoer}, Anton M. and {Pirzkal}, Nor and {Somerville}, Rachel S. and {Yung}, L.~Y. Aaron and {Amor{\'\i}n}, Ricardo O. and {Backhaus}, Bren E. and {Behroozi}, Peter and {Bisigello}, Laura and {Bromm}, Volker and {Casey}, Caitlin M. and {Ch{\'a}vez Ortiz}, {\'O}scar A. and {Cheng}, Yingjie and {Chworowsky}, Katherine and {Cleri}, Nikko J. and {Cooper}, M.~C. and {Davis}, Kelcey and {de la Vega}, Alexander and {Elbaz}, David and {Franco}, Maximilien and {Fontana}, Adriano and {Fujimoto}, Seiji and {Giavalisco}, Mauro and {Grogin}, Norman A. and {Holwerda}, Benne W. and {Huertas-Company}, Marc and {Hirschmann}, Michaela and {Iyer}, Kartheik G. and {Jogee}, Shardha and {Jung}, Intae and {Larson}, Rebecca L. and {Lucas}, Ray A. and {Mobasher}, Bahram and {Morales}, Alexa M. and {Morley}, Caroline V. and {Mukherjee}, Sagnick and {P{\'e}rez-Gonz{\'a}lez}, Pablo G. and {Ravindranath}, Swara and {Rodighiero}, Giulia and {Rowland}, Melanie J. and {Tacchella}, Sandro and {Taylor}, Anthony J. and {Trump}, Jonathan R. and {Wilkins}, Stephen M.},
        title = "{The Complete CEERS Early Universe Galaxy Sample: A Surprisingly Slow Evolution of the Space Density of Bright Galaxies at z {\ensuremath{\sim}} 8.5{\textendash}14.5}",
      journal = {\apjl},
         year = 2024,
        month = jul,
       volume = {969},
       number = {1},
          eid = {L2},
        pages = {L2},
          doi = {10.3847/2041-8213/ad4495},
archivePrefix = {arXiv},
       eprint = {2311.04279},
 primaryClass = {astro-ph.GA},
       adsurl = {https://ui.adsabs.harvard.edu/abs/2024ApJ...969L...2F}
}

@ARTICLE{fisher:2025,
       author = {{Fisher}, R. and {Bowler}, R.~A.~A. and {Stefanon}, M. and {Rowland}, L.~E. and {Algera}, H.~S.~B. and {Aravena}, M. and {Bouwens}, R. and {Dayal}, P. and {Ferrara}, A. and {Fudamoto}, Y. and {Gulis}, C. and {Hodge}, J.~A. and {Inami}, H. and {Ormerod}, K. and {Pallottini}, A. and {Phillips}, S.~G. and {Sartorio}, N.~S. and {Schouws}, S. and {Smit}, R. and {Sommovigo}, L. and {Stark}, D.~P. and {van der Werf}, P.~P.},
        title = "{REBELS-IFU: dust attenuation curves of 12 massive galaxies at z ≃ 7}",
      journal = {\mnras},
         year = 2025,
        month = may,
       volume = {539},
       number = {1},
        pages = {109-126},
          doi = {10.1093/mnras/staf485},
archivePrefix = {arXiv},
       eprint = {2501.10541},
 primaryClass = {astro-ph.GA},
       adsurl = {https://ui.adsabs.harvard.edu/abs/2025MNRAS.539..109F}
}

@ARTICLE{Fujimoto22,
       author = {{Fujimoto}, Seiji and {Finkelstein}, Steven L. and {Burgarella}, Denis and {Carilli}, Chris L. and {Buat}, V{\'e}ronique and {Casey}, Caitlin M. and {Ciesla}, Laure and {Tacchella}, Sandro and {Zavala}, Jorge A. and {Brammer}, Gabriel and {Fudamoto}, Yoshinobu and {Ouchi}, Masami and {Valentino}, Francesco and {Cooper}, M.~C. and {Dickinson}, Mark and {Franco}, Maximilien and {Giavalisco}, Mauro and {Hutchison}, Taylor A. and {Kartaltepe}, Jeyhan S. and {Koekemoer}, Anton M. and {Kojima}, Takashi and {Larson}, Rebecca L. and {Murphy}, E.~J. and {Papovich}, Casey and {P{\'e}rez-Gonz{\'a}lez}, Pablo G. and {Somerville}, Rachel S. and {Yoon}, Ilsang and {Wilkins}, Stephen M. and {Akins}, Hollis and {Amor{\'\i}n}, Ricardo O. and {Arrabal Haro}, Pablo and {Bagley}, Micaela B. and {Chworowsky}, Katherine and {Cleri}, Nikko J. and {Cooper}, Olivia R. and {Costantin}, Luca and {Daddi}, Emanuele and {Ferguson}, Henry C. and {Grogin}, Norman A. and {Jim{\'e}nez-Andrade}, E.~F. and {Juneau}, St{\'e}phanie and {Kirkpatrick}, Allison and {Kocevski}, Dale D. and {Le Bail}, Aur{\'e}lien and {Long}, Arianna and {Lucas}, Ray A. and {Magnelli}, Benjamin and {McKinney}, Jed and {Rose}, Caitlin and {Seill{\'e}}, Lise-Marie and {Simons}, Raymond C. and {Weiner}, Benjamin J. and {Yung}, L.~Y. Aaron},
        title = "{ALMA FIR View of Ultra-high-redshift Galaxy Candidates at z {\ensuremath{\sim}} 11-17: Blue Monsters or Low-z Red Interlopers?}",
      journal = {\apj},
         year = 2023,
        month = oct,
       volume = {955},
       number = {2},
          eid = {130},
        pages = {130},
          doi = {10.3847/1538-4357/aceb67},
archivePrefix = {arXiv},
       eprint = {2211.03896},
 primaryClass = {astro-ph.GA},
       adsurl = {https://ui.adsabs.harvard.edu/abs/2023ApJ...955..130F}
}

@ARTICLE{Furlanetto22,
       author = {{Furlanetto}, Steven R. and {Mirocha}, Jordan},
        title = "{On the expected purity of photometric galaxy surveys targeting the Cosmic Dawn}",
      journal = {\mnras},
         year = 2023,
        month = aug,
       volume = {523},
       number = {4},
        pages = {5274-5279},
          doi = {10.1093/mnras/stad1799},
archivePrefix = {arXiv},
       eprint = {2208.12828},
 primaryClass = {astro-ph.GA},
       adsurl = {https://ui.adsabs.harvard.edu/abs/2023MNRAS.523.5274F}
}

@ARTICLE{gelli+2020,
       author = {{Gelli}, V. and {Salvadori}, S. and {Pallottini}, A. and {Ferrara}, A.},
        title = "{The stellar populations of high-redshift dwarf galaxies}",
      journal = {\mnras},
         year = 2020,
        month = nov,
       volume = {498},
       number = {3},
        pages = {4134-4149},
          doi = {10.1093/mnras/staa2410},
archivePrefix = {arXiv},
       eprint = {2009.03912},
 primaryClass = {astro-ph.GA},
       adsurl = {https://ui.adsabs.harvard.edu/abs/2020MNRAS.498.4134G}
}

@ARTICLE{Gelli+25,
       author = {{Gelli}, Viola and {Pallottini}, Andrea and {Salvadori}, Stefania and {Ferrara}, Andrea and {Mason}, Charlotte and {Carniani}, Stefano and {Ginolfi}, Michele},
        title = "{Temporarily Quiescent Galaxies at Cosmic Dawn: Probing Bursty Star Formation}",
      journal = {\apj},
         year = 2025,
        month = may,
       volume = {985},
       number = {1},
          eid = {126},
        pages = {126},
          doi = {10.3847/1538-4357/adc722},
archivePrefix = {arXiv},
       eprint = {2501.16418},
 primaryClass = {astro-ph.GA},
       adsurl = {https://ui.adsabs.harvard.edu/abs/2025ApJ...985..126G}
}

@ARTICLE{gelli:2021,
       author = {{Gelli}, Viola and {Salvadori}, Stefania and {Ferrara}, Andrea and {Pallottini}, Andrea and {Carniani}, Stefano},
        title = "{Dwarf Satellites of High-z Lyman Break Galaxies: A Free Lunch for JWST}",
      journal = {\apjl},
         year = 2021,
        month = jun,
       volume = {913},
       number = {2},
          eid = {L25},
        pages = {L25},
          doi = {10.3847/2041-8213/abfe6c},
archivePrefix = {arXiv},
       eprint = {2105.05252},
 primaryClass = {astro-ph.GA},
       adsurl = {https://ui.adsabs.harvard.edu/abs/2021ApJ...913L..25G}
}

@ARTICLE{gelli:2023,
       author = {{Gelli}, Viola and {Salvadori}, Stefania and {Ferrara}, Andrea and {Pallottini}, Andrea and {Carniani}, Stefano},
        title = "{Quiescent Low-mass Galaxies Observed by JWST in the Epoch of Reionization}",
      journal = {\apjl},
         year = 2023,
        month = sep,
       volume = {954},
       number = {1},
          eid = {L11},
        pages = {L11},
          doi = {10.3847/2041-8213/acee80},
archivePrefix = {arXiv},
       eprint = {2303.13574},
 primaryClass = {astro-ph.GA},
       adsurl = {https://ui.adsabs.harvard.edu/abs/2023ApJ...954L..11G}
}

@ARTICLE{Grassi+14,
   author = {{Grassi}, T. and {Bovino}, S. and {Schleicher}, D.~R.~G. and 
	{Prieto}, J. and {Seifried}, D. and {Simoncini}, E. and {Gianturco}, F.~A.
	},
    title = "{KROME - a package to embed chemistry in astrophysical simulations}",
  journal = {\mnras},
     year = 2014,
    month = apr,
   volume = 439,
    pages = {2386-2419},
      doi = {10.1093/mnras/stu114},
   adsurl = {http://adsabs.harvard.edu/abs/2014MNRAS.439.2386G}
}

@ARTICLE{Grudic+21,
       author = {{Grudi{\'c}}, Michael and {Gurvich}, Alexander},
        title = "{pytreegrav: A fast Python gravity solver}",
      journal = {The Journal of Open Source Software},
         year = 2021,
        month = dec,
       volume = {6},
       number = {68},
          eid = {3675},
        pages = {3675},
          doi = {10.21105/joss.03675},
       adsurl = {https://ui.adsabs.harvard.edu/abs/2021JOSS....6.3675G}
}

@ARTICLE{Harikane+20,
       author = {{Harikane}, Yuichi and {Ouchi}, Masami and {Inoue}, Akio K. and {Matsuoka}, Yoshiki and {Tamura}, Yoichi and {Bakx}, Tom and {Fujimoto}, Seiji and {Moriwaki}, Kana and {Ono}, Yoshiaki and {Nagao}, Tohru and {Tadaki}, Ken-ichi and {Kojima}, Takashi and {Shibuya}, Takatoshi and {Egami}, Eiichi and {Ferrara}, Andrea and {Gallerani}, Simona and {Hashimoto}, Takuya and {Kohno}, Kotaro and {Matsuda}, Yuichi and {Matsuo}, Hiroshi and {Pallottini}, Andrea and {Sugahara}, Yuma and {Vallini}, Livia},
        title = "{Large Population of ALMA Galaxies at z > 6 with Very High [O III] 88 {\ensuremath{\mu}}m to [C II] 158 {\ensuremath{\mu}}m Flux Ratios: Evidence of Extremely High Ionization Parameter or PDR Deficit?}",
      journal = {\apj},
         year = 2020,
        month = jun,
       volume = {896},
       number = {2},
          eid = {93},
        pages = {93},
          doi = {10.3847/1538-4357/ab94bd},
archivePrefix = {arXiv},
       eprint = {1910.10927},
 primaryClass = {astro-ph.GA},
       adsurl = {https://ui.adsabs.harvard.edu/abs/2020ApJ...896...93H}
}

@ARTICLE{Harikane+24,
       author = {{Harikane}, Yuichi and {Inoue}, Akio K. and {Ellis}, Richard S. and {Ouchi}, Masami and {Nakazato}, Yurina and {Yoshida}, Naoki and {Ono}, Yoshiaki and {Sun}, Fengwu and {Sato}, Riku A. and {Ferrami}, Giovanni and {Fujimoto}, Seiji and {Kashikawa}, Nobunari and {McLeod}, Derek J. and {P{\'e}rez-Gonz{\'a}lez}, Pablo G. and {Sawicki}, Marcin and {Sugahara}, Yuma and {Xu}, Yi and {Yamanaka}, Satoshi and {Carnall}, Adam C. and {Cullen}, Fergus and {Dunlop}, James S. and {Egami}, Eiichi and {Grogin}, Norman and {Isobe}, Yuki and {Koekemoer}, Anton M. and {Laporte}, Nicolas and {Lee}, Chien-Hsiu and {Magee}, Dan and {Matsuo}, Hiroshi and {Matsuoka}, Yoshiki and {Mawatari}, Ken and {Nakajima}, Kimihiko and {Nakane}, Minami and {Tamura}, Yoichi and {Umeda}, Hiroya and {Yanagisawa}, Hiroto},
        title = "{JWST, ALMA, and Keck Spectroscopic Constraints on the UV Luminosity Functions at z {\ensuremath{\sim}} 7{\textendash}14: Clumpiness and Compactness of the Brightest Galaxies in the Early Universe}",
      journal = {\apj},
         year = 2025,
        month = feb,
       volume = {980},
       number = {1},
          eid = {138},
        pages = {138},
          doi = {10.3847/1538-4357/ad9b2c},
archivePrefix = {arXiv},
       eprint = {2406.18352},
 primaryClass = {astro-ph.GA},
       adsurl = {https://ui.adsabs.harvard.edu/abs/2025ApJ...980..138H}
}

@ARTICLE{Harikane+24_CEERS,
       author = {{Harikane}, Yuichi and {Nakajima}, Kimihiko and {Ouchi}, Masami and {Umeda}, Hiroya and {Isobe}, Yuki and {Ono}, Yoshiaki and {Xu}, Yi and {Zhang}, Yechi},
        title = "{Pure Spectroscopic Constraints on UV Luminosity Functions and Cosmic Star Formation History from 25 Galaxies at z $_{spec}$ = 8.61-13.20 Confirmed with JWST/NIRSpec}",
      journal = {\apj},
         year = 2024,
        month = jan,
       volume = {960},
       number = {1},
          eid = {56},
        pages = {56},
          doi = {10.3847/1538-4357/ad0b7e},
archivePrefix = {arXiv},
       eprint = {2304.06658},
 primaryClass = {astro-ph.GA},
       adsurl = {https://ui.adsabs.harvard.edu/abs/2024ApJ...960...56H}
}

@ARTICLE{Harikane22,
       author = {{Harikane}, Yuichi and {Ouchi}, Masami and {Oguri}, Masamune and {Ono}, Yoshiaki and {Nakajima}, Kimihiko and {Isobe}, Yuki and {Umeda}, Hiroya and {Mawatari}, Ken and {Zhang}, Yechi},
        title = "{A Comprehensive Study of Galaxies at z   9-16 Found in the Early JWST Data: Ultraviolet Luminosity Functions and Cosmic Star Formation History at the Pre-reionization Epoch}",
      journal = {\apjs},
         year = 2023,
        month = mar,
       volume = {265},
       number = {1},
          eid = {5},
        pages = {5},
          doi = {10.3847/1538-4365/acaaa9},
archivePrefix = {arXiv},
       eprint = {2208.01612},
 primaryClass = {astro-ph.GA},
       adsurl = {https://ui.adsabs.harvard.edu/abs/2023ApJS..265....5H}
}

@ARTICLE{Heintz+25,
       author = {{Heintz}, Kasper E. and {Pollock}, Clara L. and {Witstok}, Joris and {Carniani}, Stefano and {Hainline}, Kevin N. and {D'Eugenio}, Francesco and {Terp}, Chamilla and {Saxena}, Aayush and {Watson}, Darach},
        title = "{Dissecting the Massive Pristine, Neutral Gas Reservoir of a Remarkably Bright Galaxy at z = 14.179}",
      journal = {\apjl},
         year = 2025,
        month = jul,
       volume = {987},
       number = {1},
          eid = {L2},
        pages = {L2},
          doi = {10.3847/2041-8213/ade393},
archivePrefix = {arXiv},
       eprint = {2502.06016},
 primaryClass = {astro-ph.GA},
       adsurl = {https://ui.adsabs.harvard.edu/abs/2025ApJ...987L...2H}
}

@ARTICLE{Helton+25,
       author = {{Helton}, Jakob M. and {Rieke}, George H. and {Alberts}, Stacey and {Wu}, Zihao and {Eisenstein}, Daniel J. and {Hainline}, Kevin N. and {Carniani}, Stefano and {Ji}, Zhiyuan and {Baker}, William M. and {Bhatawdekar}, Rachana and {Bunker}, Andrew J. and {Cargile}, Phillip A. and {Charlot}, St{\'e}phane and {Chevallard}, Jacopo and {D'Eugenio}, Francesco and {Egami}, Eiichi and {Johnson}, Benjamin D. and {Jones}, Gareth C. and {Lyu}, Jianwei and {Maiolino}, Roberto and {P{\'e}rez-Gonz{\'a}lez}, Pablo G. and {Rieke}, Marcia J. and {Robertson}, Brant and {Saxena}, Aayush and {Scholtz}, Jan and {Shivaei}, Irene and {Sun}, Fengwu and {Tacchella}, Sandro and {Whitler}, Lily and {Williams}, Christina C. and {Willmer}, Christopher N.~A. and {Willott}, Chris and {Witstok}, Joris and {Zhu}, Yongda},
        title = "{Photometric detection at 7.7 {\ensuremath{\mu}}m of a galaxy beyond redshift 14 with JWST/MIRI}",
      journal = {Nature Astronomy},
         year = 2025,
        month = mar,
          doi = {10.1038/s41550-025-02503-z},
archivePrefix = {arXiv},
       eprint = {2405.18462},
 primaryClass = {astro-ph.GA},
       adsurl = {https://ui.adsabs.harvard.edu/abs/2025NatAs.tmp...66H}
}

@ARTICLE{Hsiao23,
       author = {{Hsiao}, Tiger Yu-Yang and {Abdurro'uf} and {Coe}, Dan and {Larson}, Rebecca L. and {Jung}, Intae and {Mingozzi}, Matilde and {Dayal}, Pratika and {Kumari}, Nimisha and {Kokorev}, Vasily and {Vikaeus}, Anton and {Brammer}, Gabriel and {Furtak}, Lukas J. and {Adamo}, Angela and {Andrade-Santos}, Felipe and {Antwi-Danso}, Jacqueline and {Brada{\v{c}}}, Maru{\v{s}}a and {Bradley}, Larry D. and {Broadhurst}, Tom and {Carnall}, Adam C. and {Conselice}, Christopher J. and {Diego}, Jose M. and {Donahue}, Megan and {Eldridge}, Jan J. and {Fujimoto}, Seiji and {Henry}, Alaina and {Hernandez}, Svea and {Hutchison}, Taylor A. and {James}, Bethan L. and {Norman}, Colin and {Park}, Hyunbae and {Pirzkal}, Norbert and {Postman}, Marc and {Ricotti}, Massimo and {Rigby}, Jane R. and {Vanzella}, Eros and {Welch}, Brian and {Wilkins}, Stephen M. and {Windhorst}, Rogier A. and {Xu}, Xinfeng and {Zackrisson}, Erik and {Zitrin}, Adi},
        title = "{JWST NIRSpec Spectroscopy of the Triply Lensed z = 10.17 Galaxy MACS0647{\textendash}JD}",
      journal = {\apj},
         year = 2024,
        month = sep,
       volume = {973},
       number = {1},
          eid = {8},
        pages = {8},
          doi = {10.3847/1538-4357/ad5da8},
archivePrefix = {arXiv},
       eprint = {2305.03042},
 primaryClass = {astro-ph.GA},
       adsurl = {https://ui.adsabs.harvard.edu/abs/2024ApJ...973....8H}
}

@ARTICLE{Hunter2007,
author={J. D. Hunter},
journal={Computing in Science Engineering},
title={Matplotlib: A 2D Graphics Environment},
year={2007},
volume={9},
number={3},
pages={90-95},
doi={10.1109/MCSE.2007.55},
ISSN={1521-9615},
month={May},}

@ARTICLE{Kaasinen22,
       author = {{Kaasinen}, M. and {van Marrewijk}, J. and {Popping}, G. and {Ginolfi}, M. and {Di Mascolo}, L. and {Mroczkowski}, T. and {Concas}, A. and {Di Cesare}, C. and {Killi}, M. and {Langan}, I.},
        title = "{To see or not to see a z {\ensuremath{\sim}} 13 galaxy, that is the question. Targeting the [C II] 158 {\ensuremath{\mu}}m emission line of HD1 with ALMA}",
      journal = {\aap},
         year = 2023,
        month = mar,
       volume = {671},
          eid = {A29},
        pages = {A29},
          doi = {10.1051/0004-6361/202245093},
archivePrefix = {arXiv},
       eprint = {2210.03754},
 primaryClass = {astro-ph.GA},
       adsurl = {https://ui.adsabs.harvard.edu/abs/2023A&A...671A..29K}
}

@ARTICLE{Kartaltepe+23,
       author = {{Kartaltepe}, Jeyhan S. and {Rose}, Caitlin and {Vanderhoof}, Brittany N. and {McGrath}, Elizabeth J. and {Costantin}, Luca and {Cox}, Isabella G. and {Yung}, L.~Y. Aaron and {Kocevski}, Dale D. and {Wuyts}, Stijn and {Ferguson}, Henry C. and {Bagley}, Micaela B. and {Finkelstein}, Steven L. and {Amor{\'\i}n}, Ricardo O. and {Andrews}, Brett H. and {Haro}, Pablo Arrabal and {Backhaus}, Bren E. and {Behroozi}, Peter and {Bisigello}, Laura and {Calabr{\`o}}, Antonello and {Casey}, Caitlin M. and {Coogan}, Rosemary T. and {Cooper}, M.~C. and {Croton}, Darren and {de la Vega}, Alexander and {Dickinson}, Mark and {Fontana}, Adriano and {Franco}, Maximilien and {Grazian}, Andrea and {Grogin}, Norman A. and {Hathi}, Nimish P. and {Holwerda}, Benne W. and {Huertas-Company}, Marc and {Iyer}, Kartheik G. and {Jogee}, Shardha and {Jung}, Intae and {Kewley}, Lisa J. and {Kirkpatrick}, Allison and {Koekemoer}, Anton M. and {Liu}, James and {Lotz}, Jennifer M. and {Lucas}, Ray A. and {Newman}, Jeffrey A. and {Pacifici}, Camilla and {Pandya}, Viraj and {Papovich}, Casey and {Pentericci}, Laura and {P{\'e}rez-Gonz{\'a}lez}, Pablo G. and {Petersen}, Jayse and {Pirzkal}, Nor and {Rafelski}, Marc and {Ravindranath}, Swara and {Simons}, Raymond C. and {Snyder}, Gregory F. and {Somerville}, Rachel S. and {Stanway}, Elizabeth R. and {Straughn}, Amber N. and {Tacchella}, Sandro and {Trump}, Jonathan R. and {Vega-Ferrero}, Jes{\'u}s and {Wilkins}, Stephen M. and {Yang}, Guang and {Zavala}, Jorge A.},
        title = "{CEERS Key Paper. III. The Diversity of Galaxy Structure and Morphology at z = 3-9 with JWST}",
      journal = {\apjl},
         year = 2023,
        month = mar,
       volume = {946},
       number = {1},
          eid = {L15},
        pages = {L15},
          doi = {10.3847/2041-8213/acad01},
archivePrefix = {arXiv},
       eprint = {2210.14713},
 primaryClass = {astro-ph.GA},
       adsurl = {https://ui.adsabs.harvard.edu/abs/2023ApJ...946L..15K}
}

@Article{knebe:2013,
  author    = {{Knebe}, Alexander and {Libeskind}, Noam I. and {Pearce},
          Frazer and {Behroozi}, Peter and {Casado}, Javier and
          {Dolag}, Klaus and {Dominguez-Tenreiro}, Rosa and {Elahi},
          Pascal and {Lux}, Hanni and {Muldrew}, Stuart I. and
          {Onions}, Julian},
  title     = "{Galaxies going MAD: the Galaxy-Finder Comparison
          Project}",
  journal   = {\mnras},
  year      = 2013,
  month     = jan,
  volume    = {428},
  number    = {3},
  pages     = {2039-2052},
  doi       = {10.1093/mnras/sts173},
  archiveprefix = {arXiv},
  eprint    = {1210.2578},
  primaryclass  = {astro-ph.CO},
  adsurl    = {https://ui.adsabs.harvard.edu/abs/2013MNRAS.428.2039K}
}

@ARTICLE{Kohandel+19,
       author = {{Kohandel}, M. and {Pallottini}, A. and {Ferrara}, A. and {Zanella}, A. and
         {Behrens}, C. and {Carniani}, S. and {Gallerani}, S. and {Vallini}, L.},
        title = "{Kinematics of z {\ensuremath{\geq}} 6 galaxies from [C II] line emission}",
      journal = {\mnras},
         year = "2019",
        month = "Aug",
       volume = {487},
       number = {3},
        pages = {3007-3020},
          doi = {10.1093/mnras/stz1486},
archivePrefix = {arXiv},
       eprint = {1905.11413},
 primaryClass = {astro-ph.GA},
       adsurl = {https://ui.adsabs.harvard.edu/abs/2019MNRAS.487.3007K}
}

@ARTICLE{Kohandel+20,
       author = {{Kohandel}, M. and {Pallottini}, A. and {Ferrara}, A. and {Carniani}, S. and {Gallerani}, S. and {Vallini}, L. and {Zanella}, A. and {Behrens}, C.},
        title = "{Velocity dispersion in the interstellar medium of early galaxies}",
      journal = {\mnras},
         year = 2020,
        month = nov,
       volume = {499},
       number = {1},
        pages = {1250-1265},
          doi = {10.1093/mnras/staa2792},
archivePrefix = {arXiv},
       eprint = {2009.05049},
 primaryClass = {astro-ph.GA},
       adsurl = {https://ui.adsabs.harvard.edu/abs/2020MNRAS.499.1250K}
}

@ARTICLE{Kohandel+23,
       author = {{Kohandel}, M. and {Ferrara}, A. and {Pallottini}, A. and {Vallini}, L. and {Sommovigo}, L. and {Ziparo}, F.},
        title = "{Interpreting ALMA non-detections of JWST super-early galaxies}",
      journal = {\mnras},
         year = 2023,
        month = mar,
       volume = {520},
       number = {1},
        pages = {L16-L20},
          doi = {10.1093/mnrasl/slac166},
archivePrefix = {arXiv},
       eprint = {2212.02519},
 primaryClass = {astro-ph.GA},
       adsurl = {https://ui.adsabs.harvard.edu/abs/2023MNRAS.520L..16K}
}

@ARTICLE{Kohandel+24,
       author = {{Kohandel}, M. and {Pallottini}, A. and {Ferrara}, A. and {Zanella}, A. and {Rizzo}, F. and {Carniani}, S.},
        title = "{Dynamically cold disks in the early Universe: Myth or reality?}",
      journal = {\aap},
         year = 2024,
        month = may,
       volume = {685},
          eid = {A72},
        pages = {A72},
          doi = {10.1051/0004-6361/202348209},
archivePrefix = {arXiv},
       eprint = {2311.05832},
 primaryClass = {astro-ph.GA},
       adsurl = {https://ui.adsabs.harvard.edu/abs/2024A&A...685A..72K}
}

@ARTICLE{Lazar2020,
       author = {{Lazar}, Alexandres and {Bullock}, James S. and {Boylan-Kolchin}, Michael and {Chan}, T.~K. and {Hopkins}, Philip F. and {Graus}, Andrew S. and {Wetzel}, Andrew and {El-Badry}, Kareem and {Wheeler}, Coral and {Straight}, Maria C. and {Kere{\v{s}}}, Du{\v{s}}an and {Faucher-Gigu{\`e}re}, Claude-Andr{\'e} and {Fitts}, Alex and {Garrison-Kimmel}, Shea},
        title = "{A dark matter profile to model diverse feedback-induced core sizes of {\ensuremath{\Lambda}}CDM haloes}",
      journal = {\mnras},
         year = 2020,
        month = sep,
       volume = {497},
       number = {2},
        pages = {2393-2417},
          doi = {10.1093/mnras/staa2101},
archivePrefix = {arXiv},
       eprint = {2004.10817},
 primaryClass = {astro-ph.GA},
       adsurl = {https://ui.adsabs.harvard.edu/abs/2020MNRAS.497.2393L}
}

\begin{appendix}

\section{Sub-grid model for \OIII~emission at $z > 12$}\label{sec:appendix1}

At redshifts $z > 12$, the ISM of galaxies in our simulations is not sufficiently resolved to apply radiative transfer models such as \code{Cloudy} with confidence. To provide a physically motivated estimate of \OIII~\,88$\mu$m luminosity in this regime, we adopted a sub-grid model that links unresolved star-forming activity to local gas conditions on a per-cell basis. We assigned [OIII] luminosity to each gas cell according to the following prescription:
\begin{equation}
L_{\mathrm{[OIII]88}} = \epsilon_0(n, T, Z)\, \left(\frac{f_{\mathrm{O}}}{0.7}\right)\, \left(\frac{Z}{Z_0}\right)\, \mathrm{SFR}_{\mathrm{global}} \quad [L_\odot],
\end{equation}
where $Z$ is the local gas metallicity,
$f_{\mathrm{O}} = 0.7$ is the assumed oxygen mass fraction in metals,
$Z_0 = 0.2\,Z_\odot$ is a reference metallicity,
$\mathrm{SFR}_{\mathrm{global}}$ is the total SFR within the field of view, averaged over the past 10 Myr,
 and $\epsilon_0(n, T, Z)$ is a dynamic efficiency factor that depends on local ISM conditions. The efficiency parameter is defined as\begin{equation}
\epsilon_0(n, T, Z) = \epsilon_\mathrm{base} \left( \frac{n}{n_0} \right)^\alpha \left( \frac{T}{T_0} \right)^\beta \left( \frac{Z}{Z_0} \right)^\gamma,
\end{equation}
with $\epsilon_\mathrm{base} = 3 \times 10^4\,L_\odot\,M_\odot^{-1}\,\mathrm{yr}$,
$n_0 = 10\,\mathrm{cm^{-3}}$, $T_0 = 10^4\,\mathrm{K}$, and $Z_0 = 0.2\,Z_\odot$, $\alpha = 0.5$, $\beta = -0.5$, and $\gamma = 1.0$. The model is applied only to dense, cool gas ($n > 100\,\mathrm{cm^{-3}}$ and $T < 10^4\,\mathrm{K}$), expected to dominate fine-structure line emission in unresolved ISM conditions. To avoid unphysical values from extreme fluctuations in gas properties, we capped the efficiency at $\epsilon_0 \leq 10^6\, L_\odot\, M_\odot^{-1}\,\mathrm{yr}$.

The resulting cell-by-cell luminosity is projected along the line of sight and normalized by volume to produce surface brightness maps in units of $L_\odot\,\mathrm{kpc^{-2}}$. While approximate, this model provides a physically motivated lower limit on \OIII~emission in the unresolved ISM regime.

\end{appendix}

\end{document}